\documentclass[aps,prd,preprint,groupedaddress]{revtex4-1}
\usepackage{graphicx}
\usepackage{amsmath}
\usepackage{bm}

\begin{document}
\begin{flushright}
  OU-HET-745 \ \\
\end{flushright}
\vspace{0mm}


\title{More on the relation between the two physically inequivalent
decompositions of the nucleon spin and momentum}


\author{M.~Wakamatsu}
\email[]{wakamatu@phys.sci.osaka-u.ac.jp}
\affiliation{Department of Physics, Faculty of Science, \\
Osaka University, \\
Toyonaka, Osaka 560-0043, Japan}



\begin{abstract}
In a series of papers, we have established the existence of two
gauge-invariant decompositions of the nucleon spin, which are physically
nonequivalent. The orbital angular momenta of quarks and gluons appearing
in these two decompositions are gauge-invariant dynamical
orbital angular momenta and ``generalized'' canonical orbital angular momenta 
with gauge-invariance, respectively.
The key quantity, which characterizes the difference between these two
types of orbital angular momenta is what-we-call the
{\it potential angular momentum}. We argue that the physical meaning of 
the potential angular momentum in the nucleon can
be made more transparent, by investigating a related but much simpler
example from electrodynamics. We also make clear several remaining issues in
the spin and momentum decomposition problem of the nucleon.
We clarify the relationship between the evolution equations of orbital
angular momenta corresponding to the two different decompositions above.
We also try to answer the question whether the two different decompositions of
the nucleon momentum really lead to different evolution equations, thereby
predicting conflicting asymptotic values for the quark and gluon momentum
fractions in the nucleon. 
\end{abstract}


\maketitle


\section{Introduction}
The nucleon spin puzzle raised by the EMC measurement in 1988 is
still one of the fundamental unsolved problems in QCD \cite{EMC88},\cite{EMC89}.
The current status and homework of the nucleon spin problem can
very briefly be summarized as follows. (For recent reviews, see, for
example, \cite{Rev09},\cite{Rev10}.)
First, the intrinsic quark spin contribution (or the quark polarization
in the nucleon) was fairly precisely determined to be around
1/3 \cite{COMPASS05}\nocite{COMPASS07}-\cite{HERMES07}. 
Second, gluon polarization is likely to be small,
although with large uncertainties \cite{COMPASSG06}\nocite{PHENIX06}
\nocite{STAR06A}-\cite{STAR06B}.
So, what carries the remaining 2/3 of the nucleon spin ?
That is a fundamental question of QCD which we want to answer.
To answer this question unambiguously, we cannot avoid to clarify the
following issues. What is a precise definition of each term of the
decomposition based on quantum chromodynamics (QCD) ? How can we extract
individual term by means of direct measurements? 
Let us call it the nucleon spin decomposition problem
\cite{JM90}\nocite{Ji97PRL}\nocite{BJ99}\nocite{SW00}-\cite{BLT04}.

The recent papers by Chen et al. \cite{Chen08},\cite{Chen09} arose much
controversy on the feasibility as well as the observability of the complete
decomposition of the nucleon spin \cite{Tiwari08}\nocite{Ji11A}\nocite{Ji11B}
\nocite{Wakamatsu10}\nocite{Wakamatsu11A}\nocite{Wakamatsu11B}\nocite{Cho10A}
\nocite{Cho10B}\nocite{Hatta11}\nocite{Hatta12}\nocite{Wong10}\nocite{Wang10}
\nocite{Sun10}\nocite{Chen11A}\nocite{Chen11B}\nocite{ZhangPak11}
\nocite{LinLiu11}\nocite{LinLiu12}\nocite{Chen12}\nocite{JXZ12}\nocite{JXY12}
\nocite{Burkardt12}-\cite{Lorce12}.
In the previous papers \cite{Wakamatsu10},\cite{Wakamatsu11A}, we have
established the existence of
two physically nonequivalent decompositions of the nucleon spin, both
of which are gauge-invariant. 
The quark and gluon intrinsic spin parts of these two decompositions are
nothing different in these two decompositions.
The difference appears in the orbital parts.
The quark and gluon orbital angular momenta appearing in one decomposition
is gauge-invariant {\it dynamical} (or {\it mechanical}) orbital angular
momentum (OAM), while those appearing in another decomposition is
generalized {\it canonical} OAM having gauge-invariance.
The key quantity, which characterizes the difference between these two types
of OAMs, is what-we-call the {\it potential angular momentum} \cite{Wakamatsu10}.
Understanding its physical meaning is therefore of vital importance to make
clear why there exist two decompositions at all and in what essential
respects they are different.
One of the purposes of the present paper is to clarify the physical
meaning of this potential angular momentum term in a clearest fashion with
the help of a plainer example from electrodynamics, i.e. through an analysis
of a system of charged particles and photons, analogous to a system of
color-charged quarks and gluons.

We also try to clarify several other issues left in the decomposition
problem of the nucleon spin momentum. It is known that there also exist
two different gauge-invariant decompositions of the
nucleon momentum into the contributions of quarks and gluons.
On the basis of a gauge-invariant decomposition of the nucleon momentum,
which is different from the standardly-known one, Chen et al. threw doubt
on a common wisdom of deep-inelastic-scattering physics that the gluons
carry about half of the nucleon momentum in the asymptotic limit \cite{Chen09}.
To verify the validity of this claim is of fundamental importance,
since it challenges our common knowledge on one of the basics of
perturbative QCD.   

Also important to understand is a puzzling observation on the scale dependencies
of the quark and gluon OAMs. In view of the physical inequivalence of the two
types of OAMs, i.e. the dynamical OAMs and the (generalized) canonical OAMs,
one might expect that they obey different evolution equations.
However, the past researches indicate that they do obey the same evolution
equation at least at the 1-loop level \cite{JTH96}\nocite{HJL98}-\cite{Teryaev98}.
The reason of this somewhat mysterious
observation need explanation.

The plan of the paper is as follows. To make the paper self-contained,
we briefly summarize, in sect.II, the current status of the nucleon spin
decomposition problem from our
own viewpoint. Next in sect.III, we clarify the physical meaning of
the potential angular momentum, which characterizes the difference
between the two types of OAMs, and consequently the difference between the
two inequivalent decompositions of the nucleon spin. 
Sect.V is devoted to the discussion on the relation between the two
different decompositions of the nucleon momentum. It will be shown that
the two decompositions lead to the same evolution equation at least for the
longitudinal momentum fractions of quarks and gluons, thereby predicting
the same asymptotic limits for them.
The relation between the evolution equations for the quark and gluon
OAMs corresponding to the two different decompositions of the nucleon spin
is discussed in sect.V. Then, we summarize what we have found in sect.VI.

\section{Brief review of the nucleon spin decomposition
problem \ \ \ - Where are we now ? -}
As is widely known, there have been two popular decompositions of the
nucleon spin. One is the Jaffe-Manohar decomposition \cite{JM90}, and the
other is the Ji decomposition \cite{Ji97PRL},\cite{Ji97}.
Only the intrinsic quark spin part is common in these popular
decompositions and the other parts are all different.
A disadvantage of the Jaffe-Manohar decomposition is that
each term is not separately gauge-invariant except for the
quark spin part. On the other hand, each term of the Ji decomposition is
separately gauge-invariant. Unfortunately, further gauge-invariant
decomposition of $J^g$ into its spin and orbital parts is given up in
this widely-known decomposition. Especially annoying fact was that the
sum of the gluon spin and OAM in the Jaffe-Manohar decomposition does not
coincide with the total angular momentum of gluons in the Ji decomposition.
Undoubtedly, this observation is inseparably connected with the fact that
the quark OAMs in the two decompositions are also different.

In fact, first pay attention to the difference of the quark OAM parts in the
two decompositions. What appears in the Jaffe-Manohar decomposition is
the so-called canonical OAM, which is not gauge-invariant.
On the other hand, what appears in the Ji decomposition is the so-called
dynamical (or mechanical) OAM, which is manifestly
gauge-invariant \cite{BookSakurai95}.
As is well-known, the gauge principle in physics dictates that observables
must be gauge-invariant. Because of this reason, the observability of
the canonical OAM has been questioned for a long time. 
On the other hand, Ji showed that the gauge-invariant dynamical quark OAM
can be extracted from the combined analysis of unpolarized
generalized parton distributions and the longitudinally
polarized parton distributions \cite{Ji97PRL},\cite{Ji97}.

Some years ago, however, Chen et al. proposed a new gauge-invariant
decomposition of nucleon spin \cite{Chen08},\cite{Chen09}.
The basic idea is to decompose the gluon
field $\bm{A}$ into two parts, i.e. the physical part $\bm{A}_{phys}$ and
the pure-gauge part $\bm{A}_{pure}$, which is a generalization of the
decomposition of the photon field $\bm{A}$ in QED into the transverse
component $\bm{A}_\perp$ and the longitudinal component $\bm{A}_\parallel$.
In addition to general conditions of decomposition, by imposing one plausible
theoretical constraint, Chen et al. proposed a new decomposition of the nucleon
spin. A prominent feature of their decomposition is that each term is separately
gauge-invariant, while allowing the decomposition of the total gluon
angular momentum into its spin and orbital parts.
Another noteworthy feature of this decomposition is
that it reduces to the gauge-variant decomposition of Jaffe and Manohar
in a particular gauge, $\bm{A}_{pure} = 0, \,\bm{A} = \bm{A}_{phys}$
\cite{JM90}.

Chen et al.'s papers \cite{Chen08},\cite{Chen09} arose much
controversy on the feasibility as well as the observability of the complete
decomposition of the nucleon spin \cite{Tiwari08}\nocite{Ji11A}\nocite{Ji11B}
\nocite{Wakamatsu10}\nocite{Wakamatsu11A}\nocite{Wakamatsu11B}\nocite{Cho10A}
\nocite{Cho10B}\nocite{Hatta11}\nocite{Hatta12}\nocite{Wong10}\nocite{Wang10}
\nocite{ZhangPak11}\nocite{LinLiu11}-\cite{Chen12}.
We believe that we have arrived at one satisfactory solution to the problem,
through a series of papers \cite{Wakamatsu10},\cite{Wakamatsu11A},\cite{Wakamatsu11B}.
In the 1st paper \cite{Wakamatsu10}, we have shown that the way of gauge-invariant
decomposition of nucleon spin is not necessarily unique, and proposed
another gauge-invariant decomposition.
The characteristic features of this decomposition is as follows.
First, the quark part of this decomposition is common with the Ji
decomposition, including both of spin and OAM parts.
Second, the quark and gluon spin parts are common with the Chen
decomposition. A crucial difference with the Chen decomposition appears
in the orbital parts. The sum of the quark and gluon OAMs in both
decompositions is just the same, but each term is different.
The difference of the gluon OAM in the two decompositions, which is equal
to the difference of the quark OAM in the two decompositions with an extra
minus sign, is given in the following form 
\begin{eqnarray}
  \bm{L}^g - \bm{L}^{\prime g} \ = \ - \,(\bm{L}^q - \bm{L}^{\prime q})
  \ = \ \int \,\rho^a \,(\bm{x} \times \bm{A}^a_{phys}) \,d^3 x
  \ = \ \int \, \psi^\dagger \,\bm{x} \times 
  \bm{A}_{phys} \, \psi \,\,d^3 x ,
\end{eqnarray}
and we call it the {\it potential angular momentum} by the following reason.
(In the above equation, $\bm{L}^q$ and $\bm{L}^g$ stand for the quark and gluon
OAMs in our decomposition, while $\bm{L}^{\prime q}$ and $\bm{L}^{\prime g}$
the quark and gluon OAMs in the decomposition of Chen et al.)
That is, the QED correspondent of this term is nothing but the angular
momentum carried by the electromagnetic field or potential,
which appears in the famous Feynman paradox of classical
electrodynamics \cite{Feynman65}.
An arbitrariness of the decomposition arises, because this potential
angular momentum term is {\it solely} gauge-invariant.
This means that one has a freedom to shift this potential OAM term from
the gluon OAM part to the quark OAM part in our decomposition, which in fact
leads to the quark OAM in the Chen decomposition in such a way that
\begin{eqnarray}
 &\,& \bm{L}^q
 \ + \ \mbox{\tt potential angular momentum} \nonumber \\
 &=&
 \int \,\psi^\dagger \,\bm{x} \times 
 (\bm{p} - g \,\bm{A} ) \,
 \psi \,d^3 x \ + \ g \,\int \,\psi^\dagger \,
 \bm{x} \times \bm{A}_{phys} \,\psi \,d^3 x \nonumber \\
 &=&
 \int \,\psi^\dagger \,\bm{x} \times 
 (\bm{p} - g \,\bm{A}_{pure} ) \,
 \psi \,d^3 x \ \ \ \ \ \ = \ \ \ 
 \bm{L}^{\prime q} .
\end{eqnarray}

Next, in the 2nd paper \cite{Wakamatsu11A}, we found that we can make a covariant
extension of gauge-invariant decomposition of the nucleon spin. 
Covariant generalization of the decomposition has several advantages.

\begin{itemize}
\item First, it is useful to find relations to deep-inelastic-scattering
(DIS) observables.
\item Second, it is vital to prove frame-independence of the
decomposition.
\item Third, it generalizes and unifies the previously-known nucleon spin decompositions.
\end{itemize}

Basically, we find two physically different decompositions.
The decomposition (I) contains the well-known Ji decomposition \cite{Ji97PRL},
although it also allows gauge-invariant decomposition of gluon total angular momentum
into its spin and OAM parts.
The decomposition (II) contains in it three known decomposition, i.e. those of 
Bashinsky-Jaffe \cite{BJ99}, of Chen et al. \cite{Chen08},\cite{Chen09},
and of Jaffe-Manohar \cite{JM90}, as we shall discuss below.
The basis of our treatment is a decomposition of the full gauge field into
its physical and pure-gauge parts, simiar to Chen et al.~\cite{Chen08},\cite{Chen09}.
Different from their treatment, however, we impose only the following
quite general conditions :  
\begin{eqnarray}
 F^{\mu \nu}_{pure} \ \equiv \ \partial^\mu \,A^\nu_{pure} \ - \ 
 \partial^\nu \,A^\mu_{pure} \ - \ i \,g \,[A^\mu_{pure}, A^\nu_{pure}]
 \ = \ 0, \label{gcond1}
\end{eqnarray}
and
\begin{eqnarray}
 A^\mu_{phys} (x) &\rightarrow& U(x) \,A^\mu_{phys} (x) \,U^\dagger (x), \label{gcond2} \\
 A^\mu_{pure} (x) &\rightarrow& U(x) \,\left(\,A^\mu_{pure} (x) \ + \ 
 \frac{i}{g} \,\,\partial^\mu \,\,\right) \,U^\dagger (x) . \label{gcond3}
\end{eqnarray}
The first is the pure-gauge condition for $A^\mu_{pure}$, while the second
are the gauge transformation properties for these two components.
(These transformation properties indicates that the physics is basically contained
in the physical part $A^\mu_{phys}$, while the pure-gauge part $A^\mu_{pure}$
carries unphysical gauge degrees of freedom.)
Actually, these conditions are not enough to fix gauge uniquely.
However, the point of our theoretical scheme is that we can {\it postpone}
a complete gauge fixing until later stage, while accomplishing a
gauge-invariant decomposition of $M^{\mu \nu \lambda}$ based on the
above conditions alone. Still, we find the way of gauge-invariant
decomposition is not unique and are left with two possibilities.

We start with the decomposition (II) given in the form : 
\begin{equation}
 M^{\mu \nu \lambda} \ = \ M^{\prime \mu \nu \lambda}_{q-spin} \ + \ 
 M^{\prime \mu \nu \lambda}_{q-OAM} \ + \ M^{\prime \mu \nu \lambda}_{G-spin}
 \ + \ M^{\prime \mu \nu \lambda}_{G-OAM} \ + \ M^{\prime \mu \nu \lambda}_{boost},
 \label{decompositionII}
\end{equation}
with
\begin{eqnarray}
 M^{\prime \mu \nu \lambda}_{q-spin}
 &=& M^{\mu \nu \lambda}_{q-spin}, \\
 M^{\prime \mu \nu \lambda}_{q-OAM}
 &=& \bar{\psi} \,\gamma^\mu \,(\,x^{\nu} \,i \,D^{\lambda}_{pure} 
 \ - \ x^{\lambda} \,i \,D^{\nu}_{pure} \,) \,\psi \label{decomposition2B} \\
 M^{\prime \mu \nu \lambda}_{G-spin} 
 &=& M^{\mu \nu \lambda}_{G-spin}, \\
 M^{\prime \mu \nu \lambda}_{G-OAM} 
 &=& - \,2 \,\mbox{Tr} \,[\, F^{\mu \alpha} \,
 (\,x^{\nu} \,D^{\lambda}_{pure}
 \ - \ x^{\lambda} \,D^{\nu}_{pure} \,) \,A_{\alpha}^{phys} \,] ,
 \label{decomposition2E}
\end{eqnarray}
and
\begin{equation}
 M^{\prime \mu \nu \lambda}_{boost} 
 \ = \ - \,\frac{1}{2} \,\mbox{Tr} \,F^2 \,(\,x^{\nu} \,g^{\mu \lambda} 
 \ - \ x^{\lambda} \,g^{\mu \nu} \,) .
\end{equation}

At first sight, this decomposition looks like a covariant generalization
of Chen et al's decomposition, in the sense that the quark OAM part contains
{\it pure gauge covariant derivative}.
However, a crucial difference is that we have not yet fixed the gauge (and the
Lorentz frame) explicitly.
The point is that, as long as the general conditions (\ref{gcond1}),
(\ref{gcond2}), (\ref{gcond3}) are satisfied, each term of the decomposition (II)
is never mixed up under general color gauge transformation of QCD, which means
that each term is separately gauge-invariant \cite{Wakamatsu11A}.
These conditions are general enough,
so that they are expected to be satisfied by most gauges used in QCD.
The fact that the Bashinsky-Jaffe decomposition is contained in our more general
decomposition (II) was explicitly verified in \cite{Wakamatsu11A}. It is also
logically plausible that the Chen el al. decomposition is contained in our decomposition
(II). This is because, although their decomposition is given in a noncovariant
manner like the formulation of electrodynamics to be discussed in sec.II, their
decomposition of the gluon field $\bm{A}$ into $\bm{A}_{phys}$ and $\bm{A}_{pure}$
naturally satisfy our general conditions (\ref{gcond1}),(\ref{gcond2}),(\ref{gcond3}). 
In view of the fact that both frameworks of Bashinsky-Jaffe and of Chen et al.
are contained in our more general gauge-invariant decomposition, we naturally
expect that both give the same answer at least for the momentum sum rule of QCD
as well as for the longitudinal spin decomposition of the nucleon, which can be
formulated frame-independently.

In a recent paper, Ji, Xu, and Zhao threw doubt on this viewpoint \cite{JXZ12}.
According to them, the Bashinsky-Jaffe decomposition is one gauge-invariant extension
(GIE) of gauge-variant Jaffe-Manohar decomposition based on the light-cone gauge,
whereas the Chen el al's decomposition is another GIE based on
the Coulomb gauge. (Concerning the idea of gauge-invariant extension (GIE),
see also \cite{JXY12}.)
Their claim is that, since they are different GIEs, there is no reason to expect
that they give the same physical predictions.
However, it seems to us that their conclusion is heavily influenced by the following
observation. That is, the explicit calculations of the evolution matrices for the
momentum fractions of quarks and gluons by Chen et al. based on the generalized
Coulomb gauge are advertized to give totally different answers from the
standardly-believed ones, which the treatment in the light-cone gauge can reproduce
as we shall see later.
However, no one has checked the validity of their Coulomb gauge calculation yet.
What would one conclude, if this discrepancy simply arises from some technical
mistakes in the Coulomb gauge treatment of the problem ? We shall come back to
this question at the end of sect. IV.
  
Next we turn to the decomposition (I) given in the form : 
\begin{equation}
 M^{\mu \nu \lambda} \ = \ M^{\mu \nu \lambda}_{q-spin} \ + \ 
 M^{\mu \nu \lambda}_{q-OAM} \ + \ M^{\mu \nu \lambda}_{G-spin}
 \ + \ M^{\mu \nu \lambda}_{G-OAM} \ + \ M^{\mu \nu \lambda}_{boost},
 \label{decompositionI}
\end{equation}
with
\begin{eqnarray}
 M^{\mu \nu \lambda}_{q-spin}
 &=& \frac{1}{2} \,\epsilon^{\mu \nu \lambda \sigma} \,\bar{\psi} 
 \,\gamma_{\sigma} \,\gamma_5 \,\psi , \label{decomposition1A} \\
 M^{\mu \nu \lambda}_{q-OAM}
 &=& \bar{\psi} \,\gamma^\mu \,(\,x^{\nu} \,i \,D^{\lambda} 
 \ - \ x^{\lambda} \,i \,D^{\nu} \,) \,\psi \label{decomposition1B} \\
 M^{\mu \nu \lambda}_{G-spin} 
 &=& 2 \,\mbox{Tr} \,[\, F^{\mu \lambda} \,A^{\nu}_{phys} 
 \ - \ F^{\mu \nu} \,A^{\lambda}_{phys} \,], \label{decomposition1C} \\
 M^{\mu \nu \lambda}_{G-OAM} 
 &=& - \,2 \,\mbox{Tr} \,[\, F^{\mu \alpha} \,
 (\,x^{\nu} \,D^{\lambda}_{pure}
 \ - \ x^{\lambda} \,D^{\nu}_{pure} \,) \,A_{\alpha}^{phys} \,],
 \nonumber \\
 &\,& + \, 2 \,\mbox{Tr} \,[\, (\,D_{\alpha} \,F^{\alpha \mu} \,)
 \,(\,x^{\nu} \,A^{\lambda}_{phys} \ - \ 
 x^{\lambda} \,A^{\nu}_{phys} \,) \,], \\
 M^{\mu \nu \lambda}_{boost} &=& M^{\prime \mu \nu \lambda} . 
 \label{decomposition1D}
\end{eqnarray}
It differs from the decomposition (II) in the orbital parts.
The quark OAM part contains {\it full covariant derivative} contrary to the
decomposition (II). Correspondingly, the gluon OAM part is also different.
It contains a covariant generalization of the potential angular momentum
term.

It was sometimes criticized that there are so many decompositions
of the nucleon spin. As already explained, we do not take this viewpoint.
We claim that there are only two physically nonequivalent decompositions.
(We shall develop an argument which gives a support to this viewpoint,
in the next section, by utilizing a plainer example from electrodynamics.)
One is an extension of the Ji decomposition, which also fulfills the
decomposition of the gluon total angular momentum into the intrinsic spin
and orbital part, while the other is a decomposition that contains in it
three known decompositions as {\it gauge-fixed forms} of more general
expression. (We however recall that there a criticism to this idea \cite{JXZ12}.) 
The orbital OAMs appearing in these
two decompositions are respectively the {\it dynamical} OAMs and the
{\it generalized canonical} OAMs.
Since both decompositions are gauge-invariant, there arises a possibility
that they both correspond to observables.

A clear relation with observables was first obtained for
the decomposition (I)  \cite{Wakamatsu11A}.
The keys are the following identities, which hold in our decomposition (I).
For the quark part, it holds that
\begin{equation}
 x^\nu \,T^{\mu \lambda}_q \ - \ x^\lambda \,T^{\mu \nu}_q \ = \ 
 M^{\mu \nu \lambda}_{q-spin} \ + \ M^{\mu \nu \lambda}_{q-OAM} \ + \
 \mbox{total divergence} , 
\end{equation}
while for the gluon part we have
\begin{equation}
 x^\nu \,T^{\mu \lambda}_g \ - \ x^\lambda \,T^{\mu \nu}_q \ - \ 
 \mbox{boost} \ = \ 
 M^{\mu \nu \lambda}_{g-spin} \ + \ M^{\mu \nu \lambda}_{g-OAM} \ + \
 \mbox{total divergence}. 
\end{equation}
Here, $T^{\mu \nu}_q$ and $T^{\mu \nu}_g$ respectively stand for
the quark and gluon parts of QCD energy-momentum tensor in the
Belinfante symmetrized form.
By evaluating the nucleon forward matrix element of the above identities,
we can prove the following important relations.

First, for the quark part, we get 
\begin{eqnarray}
 \hspace{10mm} L_q &\equiv& 
 \langle p \uparrow \,| \,
 M^{012}_{q-OAM} \,| \, p \uparrow \rangle \nonumber \\
 &=& \frac{1}{2} \,\int_{-1}^1 \,x \,
 [\,H^q (x,0,0) \ + E^q (x,0,0) \,] \,dx \ - \ 
 \frac{1}{2} \,\int_{-1}^1 \,\Delta q(x) \,dx , 
\end{eqnarray}
with
\begin{eqnarray}
 M^{012}_{q-OAM} 
 \ = \ \bar{\psi} \,\left(\mbox{\boldmath $x$} \times \frac{1}{i} \,
 \mbox{\boldmath $D$} \right)^3 \,\psi
 &\neq& \left\{ \,
 \begin{array}{l}
 \bar{\psi} \,\left(\mbox{\boldmath $x$} \times \frac{1}{i} \,
 \nabla \right)^3
 \,\psi \\ 
 \bar{\psi} \,\left(\mbox{\boldmath $x$} \times \frac{1}{i} \,
 \mbox{\boldmath $D$}_{pure} \right)^3 \,\psi .
 \end{array}
 \right.
\end{eqnarray}
We find that the proton matrix element of our quark OAM
operator coincides with the difference between the 2nd moment of
GPD $H + E$ and the 1st moment of the longitudinally polarized distribution
of quarks.
What should be emphasized here is that full covariant derivative appears,
not a simple derivative operator nor pure-gauge covariant derivative.
In other words, the quark OAM extracted from the combined
analysis of GPD and polarized PDF is {\it dynamical} (or 
{\it mechanical}) OAM not {\it canonical} OAM.
This conclusion is nothing different from Ji's finding \cite{Ji97PRL}. 

Also for the gluon part, we find that the difference between the
2nd moment of gluon GPD $H + E$ and the 1st moment of polarized gluon
distribution coincides with the proton matrix element of our gluon OAM
operator given as follows.
\begin{eqnarray}
 \hspace{10mm} L_g &\equiv& 
 \langle p \uparrow \,| \, M^{012}_{g-OAM} \,| \,
 p \uparrow \rangle \nonumber \\
 &=& \frac{1}{2} \,\int_{-1}^1 \,x \,[\,
 H^g (x,0,0) \ + \ E^g (x,0,0) \,] \,dx \ - \ 
 \int_0^1 \,\Delta g(x) \,dx , 
\end{eqnarray}
with
\begin{eqnarray}
 \hspace{10mm} M^{012}_{g-OAM} &=& 
 2 \,\mbox{\rm Tr} \,[\,E^j \,(\mbox{\boldmath $x$} \times 
 \mbox{\boldmath $D$}_{pure})^3 \,A^{phys}_j \,] 
 \hspace{10mm} : \ \ \ 
 \mbox{\rm canonical OAM}
 \nonumber \\
 &+& 2 \,\mbox{\rm Tr} \,[\,\rho \,(\mbox{\boldmath $x$} \times
 \mbox{\boldmath $A$}_{phys})^3 \,]
 \hspace{22mm} : \ \ \ 
 \mbox{\rm potential OAM term} .
\end{eqnarray}
Namely, the gluon OAM extracted from the combined analysis of GPD and
polarized PDF contains {\it potential} OAM term, in addition to
{\it canonical} OAM. (Notice that this also clarifies the reason why
the sum of the gluon spin and OAM in the Jaffe-Manohar decomposition
does not coincide with the total gluon angular momentum in the Ji
decompostion.) It would be legitimate to call the whole part the
gluon {\it dynamical} or {\it mechanical} OAM.

Here, we want to make several important remarks on the above sum rules.
First, our decomposition has a Lorentz-frame-independent meaning.
This should be clear from the fact that the GPDs and PDFs appearing
in our sum rules are manifestly Lorentz-invariant quantities.
Recently, Goldman argued that the nucleon spin decomposition
is frame-dependent \cite{Goldman11}. This is generally true.
In fact, Leader recently proposed a sum rule for the transverse angular
momentum \cite{Leader11}. In this sum rule, $P_0$, the energy of the nucleon,
appears. It is clear that this sum rule is manifestly frame-dependent.
Note, however, that our main interest here is the simplest and most
fundamental longitudinal spin decomposition of the nucleon.
We emphasize once again that the
longitudinally spin decomposition is definitely frame-independent.
We think it a welcome feature, since, then,  the decomposition can be
thought to reflect intrinsic properties of the nucleon, which are
independent of the velocity which the nucleon is running with.
Underlying reason why the longitudinal spin sum rule is Lorentz-frame
independent seems very simple. The OAM component along the longitudinal
direction comes from the motion in the perpendicular plane to this axis,
and such transverse motion is not affected by the Lorentz transformation
along this axis.

Although our decomposition looks quite satisfactory in many respects,
one subtle question remained. It is a role of quantum-loop effects.
Is the longitudinal gluon polarization $\Delta G$ gauge-invariant even
at quantum level ? This is a fairly delicate question. 
In fact, despite the existence of several formal proof showing the
gauge-invariance of $\Delta G$ \cite{Manohar90}\nocite{Manohar91A}
\nocite{Manohar91B}\nocite{BB91}-\cite{AEL95}, it was sometimes claimed that
$\Delta G$ has its meaning only in the light-cone gauge and infinite-momentum
frame \cite{JTH96},\cite{HJL99}.
More specifically, in an influential paper, Hoodbhoy, Ji, and Lu claim
that $\Delta G$ evolves differently in the Feynman gauge and the
LC gauge \cite{HJL99}.
However, the gluon spin operator used in their Feynman gauge calculation is
given by
\begin{eqnarray}
 M^{+12}_{g-spin} &=& 2 \,\mbox{\rm Tr} \,
 [\,F^{+1} \, A^2 
 \ - \ F^{+2} \,A^1 
 \,] ,
\end{eqnarray}
which is not gauge-invariant, and is delicately different from our
gauge-invariant gluon-spin operator given as
\begin{eqnarray}
 M^{+12}_{g-spin} &=& 2 \,\mbox{\rm Tr} \,
 [\,F^{+1} \, A^2_{phys} 
 \ - \ F^{+2} \,A^1_{phys} 
 \,] ,
\end{eqnarray}
The problem is how to incorporate this difference into the Feynman rule for
evaluating 1-loop anomalous dimension of the quark and gluon spin operators.
This problem was attacked and solved in the paper \cite{Wakamatsu11B}.
We find that the calculation in the Feynman gauge (as well as in any
covariant gauge including the Landau gauge) reproduces the answer obtained
in the LC gauge, which is also the answer obtained in the famous
Altarelli-Parisi method \cite{AP77}. (This conclusion for the evolution
of $\Delta G$ however contradicts the one given in \cite{Chen11B}.)
Our finding is important also from another context. 
So far, a direct check of the answer of Altarelli-Pasiri method for the
evolution of $\Delta G$ within the operator-product-expansion (OPE)
framework was limited to the LC gauge, because it was believed that there
is no gauge-invariant definition of gluon spin in the OPE framework.
This was the reason why the question of gauge-invariance of $\Delta G$ has
been left in unclear status for a long time.

After establishing satisfactory natures of the decomposition (I),
now we turn our attention to another decomposition (II).
According to Chen et al., the greatest advantage of the
decomposition (II) is that their quark OAM operator $\bm{L}^\prime_q$
satisfies the standard commutation relation of angular momentum : 
\begin{equation}
 \bm{L}^\prime_q \times \bm{L}^\prime_q \ = \ i \,\bm{L}^\prime_q ,
\end{equation}
due to the property $\nabla \times \bm{A}_{pure} = 0$.
This property was claimed to be essential for its physical interpretation
as an OAM. However, this is not necessarily true as is clear
from the papers \cite{VanEN94A},\cite{VanEN94B}, which treats a similar
problem in QED.
It was shown there that the spin and OAM operators
of the photons do not satisfy the ordinary
commutation relation of angular momentum ($SU(2)$ algebra) separately.
This is not surprising at all. In fact, it is true that the total momentum
as well as the total angular momentum operators of a composite system must
satisfy the Poincare algebra, because, in quantum field theory, a physical
state of a composite particle must be one of the irreducible representations
of the Poincare group. However, it is not an absolute demand of the Poincare
symmetry that the momentum and the angular momentum of each constitute of a
composite particle satisfies the Poincare algebra separately.  
Then, the claimed superiority of the decomposition (II) over (I)
is not actually present.
Nevertheless, since the decomposition (II) is also gauge-invariant,
there still remains a possibility that it can be related to observables.

Recently, Hatta made an important step toward this direction \cite{Hatta12}
based on his decomposition formula of the physical- and pure-gauge
components of gluon fields proposed by himself in \cite{Hatta11}.
Starting from the gauge-invariant expression of the Wigner distribution
also called the generalized transverse-momentum-dependent distributions
(GTMD), which depends not only the longitudinal
and transverse momenta but also the momentum transfer of the target nucleon,
he showed that the nucleon matrix element of the {\it generalized canonical}
OAM can be related to a weighted integral of a certain GTMD.
It is important to recognize that this quantity does not appear in the
standard classification of TMDs by the following reason.
To explain it, we first recall the definition of the most fundamental
GTMD appearing in the classification given in \cite{MMS09} :  
\begin{eqnarray}
 &\,& W^{[\gamma^+]} (x, \xi, \bm{q}^2_T, 
 \bm{q}_T \cdot \bm{\Delta}_T,\bm{\Delta}^2_T ; \eta) \nonumber \\
 &=& \frac{1}{2} \,\int \,\frac{d z^- \,d^2 z_T}{(2 \,\pi)^2} \,
 e^{k \cdot z} \,\langle p^\prime, \lambda^\prime \,| \,
 \bar{\psi} \left(- \,\frac{z}{2} \right) \,\gamma^+ \,
 {\cal W} \left(- \,\frac{z}{2}, \frac{z}{2} \,| \,n \right) \,
 \psi \left( \frac{z}{2} \right) \,| \,p, \lambda 
 \rangle_{z^+ = 0} \nonumber \\
 &=& \frac{1}{2 \,M} \,\bar{u} (p^\prime, \lambda^\prime) \,
 \left[\,F_{1,1} \, + \, 
 \frac{i \,\sigma^{i +} \,q^i_T}{P^+} \,
 F_{1,2} \, + \, 
 \frac{i \,\sigma^{i +} \,\Delta^i_T}{P^+} \,
 F_{1,3} \, + \, 
 \frac{i \,\sigma^{ij} \,q^i_T \,\Delta^j_T}{M^2} \,\,
 F_{1,4} \,\right] \,u(p, \lambda). \hspace{10mm}
\end{eqnarray}
The GTMD defined by the 2nd line of the above equation generally
contains 4 pieces of invariant functions 
$F_{1,i} (x, \xi, \bm{q}_T^2, \bm{\Delta}_T^2, \eta)$ with $i = 1, \cdots, 4$,
which are functions of the Bjorken variable $x$, the skewedness parameter
$\xi$, the transverse-momentum square $\bm{q}_T^2$, the 
transverse-momentum-transfer square $\bm{\Delta}_T^2$, and the parameter
$\eta$ characterizing the nature of the functions under time-reversal.
In the forward limit, the first and the second pieces respectively reduce
to the usual spin-independent TMD and the naively-time-reversal-odd Sivers
function. On the other hand, the last two terms disappear in the forward limit,
$\bm{\Delta}_\perp \rightarrow 0$.
Nonetheless, within the framework of a quark model, which does not pay much
attention to the gauge-invariance issue, 
Lorce and Pasquini showed \cite{LP11} that a weighted integral of this 4th
function is related to the nucleon matrix element of the canonical OAM given as
\begin{eqnarray}
 L_{can}
 \ = \ - \,\int \, dx \,d^2 q_T \,\,
 \frac{\bm{q}^2_T}{M^2} \,\,\,F^q_{1,4}
 (x, 0, q^2_T,0,0) .
\end{eqnarray}
This is just the sum rule, to which Hatta gave a gauge-invariant meaning,
i.e. the meaning within the framework of QCD as a color gauge theory.
In this sense, Hatta's work opened up a possibility that the
OAM appearing in the decomposition (II) may also be related to observables.
Since the relation between the OAM appearing in the decomposition (I) and the observables is already known, this means that we may be able to isolate the correspondent of {\it potential angular momentum} term appearing in Feynman's 
paradox as a difference between the two OAMs as
\begin{eqnarray}
 L_{pot} \ = \ L_{mech} \ - \ L_{``can"}
\end{eqnarray}
However, one must be careful about the presence of very delicate problem on the sum rules containing GTMDs and/or ordinary TMDs. (See, for example, the textbook \cite{BookCollins},
which discusses the delicacies of TMDs in full detail.)
Once quantum loop effects are taken into account, the very existence of TMDs
satisfying gauge-invariance and factorization (universality or process independence)
simultaneously is being questioned.
Is process-independent extraction of $L_{``can"}$ really possible ?
One must say that it is still a challenging open question.

\section{What is ``potential angular momentum" ?}

We have shown that the key quantity, which distinguishes the two physically
different nucleon spin decompositions, is what-we-call the
``potential angular momentum" term.
To understand its physical meaning more throughly, and also to understand the
reason why there exist two physically different decompositions with
gauge-invariance, we find it very instructive to
study easier QED case, especially an interacting system of charged particles
and photons \cite{BookJR76}\nocite{BookBLP82}-\cite{BookCDG89}.
The total Hamiltonian of such system is given by
\begin{eqnarray}
 H &=& \sum_i \,\frac{1}{2} \,m_i \,\dot{\bm{r}}_i^2 \ + \ 
 \frac{1}{2} \,\int \,d^3 r \,[\,\bm{E}^2 + \bm{B}^2 \,] .
 \label{QED_Hamiltonian}
\end{eqnarray}
Here the 1st and the 2nd terms of the r.h.s. respectively stand for
the mechanical kinetic energy of the charged particles and the
total energy of the electromagnetic fields.
As is well-known, the vector potential $\bm{A} (\bm{r},t)$ of the photon
can be decomposed into longitudinal and transverse components as
\begin{equation}
 \bm{A} \ = \ \bm{A}_\parallel \ + \ \bm{A}_\perp, \label{long-trans_decomp}
\end{equation}
with the properties
\begin{equation}
 \nabla \times \bm{A}_\parallel \ = \ 0, \ \ \ \ 
 \nabla \cdot \bm{A}_\perp \ = \ 0 .
\end{equation}
We emphasize that this longitudinal-transverse decomposition is {\it unique},
once the Lorentz frame of reference is fixed. 
Under a general gauge transformation given as
\begin{eqnarray}
 A^0 &\rightarrow& A^{\prime 0} \ = \ A^0 \ - \ 
 \frac{\partial}{\partial t} \,\Lambda (x), \\
 \bm{A} &\rightarrow& \bm{A}^\prime \ = \ \bm{A} \ + \ 
 \nabla \Lambda (x),
\end{eqnarray}
the longitudinal component $\bm{A}_\parallel$ transforms
as
\begin{eqnarray}
 \bm{A}_\parallel &\rightarrow& \bm{A}_\parallel^\prime \ = \
 \bm{A}_\parallel \ + \ \nabla \,\Lambda (x),
\end{eqnarray}
while
the transverse component $\bm{A}_\perp$ is invariant, i.e.
\begin{equation}
 \bm{A}_\perp \ \rightarrow \ \bm{A}_\perp^\prime \ = \ 
 \bm{A}_\perp ,
\end{equation}
indicating that $\bm{A}_\parallel$ carries unphysical gauge degrees of freedom.

To avoid misunderstanding, we think it important to clarify the fact that the
decomposition (\ref{long-trans_decomp}) itself has nothing to do with
Coulomb-gauge fixing. The Coulomb-gauge condition is to require that
\begin{equation}
 \nabla \cdot \bm{A} \ = \ 0.
\end{equation}
Since $\nabla \cdot \bm{A}_\perp = 0$ by definition, this is equivalent to
requiring that
\begin{equation}
 \nabla \cdot \bm{A}_\parallel = 0.
\end{equation}
This is the Coulomb-gauge fixing condition, which works to eliminate unphysical
gauge degrees of freedom $\bm{A}_\parallel$. In fact, once this condition is
imposed, $\bm{A}_\parallel$ is divergence-free as well as irrotational,
so that we can take
\begin{equation}
 \bm{A}_\parallel = 0 .
\end{equation}
without loss of generality.

Naturally, the separation of the vector potential $\bm{A}$ of the photon
into the transverse (physical) and longitudinal (pure-gauge) components
is {\it frame-dependent}.
However, it is also true that we can start this
decomposition in an arbitrary Lorentz frame. The Coulomb gauge condition
$\nabla \cdot \bm{A} = 0$ is definitely Lorentz non-covariant, different
from the so-called Lorenz gauge condition $\partial_\mu A^\mu = 0$.
It is known that the 4-vector potential $A^\mu$ in the Lorenz gauge
satisfies the Lorenz gauge condition $\partial_\mu A^\mu = 0$ even
after Lorentz transformation to another frame. On the other hand, when
the 3-vector potential $\bm{A}$ in a certain Lorentz frame is prepared
to satisfy the Coulomb gauge condition $\nabla \cdot \bm{A} = 0$, the
Lorentz-transformed vector potential $\bm{A}^\prime$ does not satisfy
$\nabla^\prime \cdot \bm{A}^\prime = 0$. Here, we need a further gauge
transformation in order to get $\bm{A}^\prime$ satisfying
$\nabla^\prime \cdot \bm{A}^\prime = 0$.
However, this does not make any trouble because we could start the whole
consideration in the transformed frame and could impose the condition
$\nabla \cdot \bm{A} = 0$ in that frame. 
(An equivalent but formally more convenient framework for showing the
covariance of the Coulomb gauge treatment would be to require somewhat
nonstandard Lorentz transformation property for the four-vector potential
of the gauge field as described in the textbook of Bjorken and
Drell \cite{BookBD65} as well as in the recent paper \cite{Lorce12}.)
The fact is that observables
(which must of course be gauge-invariant) are independent of the choice
of gauge. Both of the Lorentz gauge and the Coulomb gauge give exactly the
same answer for physical observables.
The is just the core of Maxwell's electrodynamics as a Lorentz-invariant
gauge theory.

To return to our main discussion, in parallel with the above decomposition
of the vector potential $\bm{A}$, the electric field can also be decomposed
into longitudinal and transverse components as
\begin{eqnarray}
 \bm{E} \ = \ \bm{E}_\parallel \ + \ \bm{E}_\perp,
\end{eqnarray}
with
\begin{eqnarray}
 \bm{E}_\parallel &=& - \,\nabla A^0 \ - \ 
 \frac{\partial \bm{A}_\parallel}{\partial t}, \\
 \bm{E}_\perp &=& - \,\frac{\partial \bm{A}_\perp}{\partial t},
\end{eqnarray}
while the magnetic field is intrinsically transverse
\begin{equation}
 \bm{B} \ = \ \nabla \times \bm{A} \ = \ \nabla \times \bm{A}_\perp
 \ = \ \bm{B}_\perp.
\end{equation}
As a consequence, the photon part of the total energy can be
decomposed into two pieces, i.e. the longitudinal part and the
transverse part, as
\begin{eqnarray}
 H &=& \sum_{i-1}^N \,\frac{1}{2} \,m_i \,\dot{\bm{r}}_i^2 \ + \ 
 \frac{1}{2} \,\int \,d^3 r \,
 \bm{E}_\parallel^2 \ + \ 
 \frac{1}{2} \,\int \,d^3 r \,[\,\bm{E}_\perp^2 + \bm{B}_\perp^2 \,].
\end{eqnarray}
Now, by using the Gauss law $\nabla \cdot \bm{E}_\parallel = \rho$, 
it can be shown that the longitudinal part is nothing but the
Coulomb energy between the charged particles (aside from
the self-energies), so that we can write as
\begin{equation}
 H \ = \  
 \sum_{i = 1}^N \,\frac{1}{2} \,m_i \,\dot{\bm{r}}_i^2 \ \ + \ \ 
 V_{coul} \ \ + \ \ H_{trans} ,
\end{equation}
with
\begin{eqnarray}
 V_{Coul} &=& \frac{1}{4 \,\pi} \,\sum_{i,j = 1 \,(i \neq j)}^N \,
 \frac{q_i \,q_j}{| \bm{r}_i - \bm{r}_j |} , \\
 H_{trans} &=& \frac{1}{2} \,\int \,d^3 r \,[\,
 \bm{E}_\perp^2 + \bm{B}_\perp^2 \,] .
\end{eqnarray}

Next we consider a similar decomposition of the total momentum.
The total momentum of the system is a sum of the mechanical momentum
of charged particles and the momentum of photon fields as
\begin{equation}
 \bm{P} \ = \ \sum_i \,m_i \,\dot{\bm{r}}_i \ + \ 
 \int \,d^3 r \,\,\bm{E} \times \bm{B}.
\end{equation}
The total momentum of the electromagnetic fields can be decomposed
into longitudinal and transverse parts as
\begin{equation}
 \int \,d^3 r \,\bm{E} \times \bm{B} \ = \ \bm{P}_{long} \ + \ \bm{P}_{trans}.
\end{equation}
with
\begin{eqnarray}
 \bm{P}_{long} &=& \int \,d^3 r \, \bm{E}_\parallel \times \bm{B}_\perp, \\
 \bm{P}_{trans} &=& \int \,d^3 r \,\bm{E}_\perp \times \bm{B}_\perp,
\end{eqnarray}
which gives the decomposition
\begin{equation}
 \bm{P} \ = \ \sum_i \,m_i \,\dot{\bm{r}}_i \ + \ 
 \bm{P}_{long} \ + \ \bm{P}_{trans} .
\end{equation}
Again, by using the Gauss law, it can be shown that
$\bm{P}_{long}$ is also expressed as
\begin{equation}
 \bm{P}_{long} \ = \ \sum_i \,q_i \,\bm{A}_\perp (\bm{r}_i),
\end{equation}
so that we can write as
\begin{equation}
 \bm{P} \ = \ \sum_i \,m_i \,\dot{\bm{r}}_i \ + \ 
 \sum_i \,q_i \,\bm{A}_\perp (\bm{r}_i) \ + \ \bm{P}_{trans}.
\end{equation}
We point out that the quantity $q_i \,\bm{A}_\perp (\bm{r}_i)$ appearing
in this decomposition is nothing but the {\it potential momentum}
according to the terminology of Konopinski \cite{Konopinski78}.
In the present context, it represents the momentum that associates with the
longitudinal (electric) field generated by the particle $i$.
Which of particles or photons should it be attributed to ?
This is a fairly delicate question. It is of the same sort of question as
which of charged particles or photons should the Coulomb energy be
attributed to. To attribute it to charged particle is closer to the concept
of ``action at a distance theory", while to attribute it to electromagnetic
field is closer to the concept of ``action through medium".
If there is no difference between their physical predictions, the choice
is a matter of convenience.
Let us see what happens if we combine the potential momentum term with
the mechanical energy of charged particles.
To this end, we recall that, under the presence of electromagnetic
potential, the canonical momentum $\bm{p}_i$ of the
charged particle $i$ is given by the equation
\begin{equation}
 \bm{p}_i \ \equiv \ \frac{\partial L}{\partial \dot{\bm{r}}_i}
 \ = \ m_i \,\dot{\bm{r}}_i \ + \ q_i \,\bm{A} (\bm{r}_i) ,
\end{equation}
where $L$ is the lagrangian corresponding to the Hamiltonian (\ref{QED_Hamiltonian}). 
Using it, the total momentum $\bm{P}$ can be expressed in the following form :
\begin{equation}
 \bm{P} \ = \ \sum_i \,\left(\bm{p}_i \ - \ q_i \,
 \bm{A}_\parallel (\bm{r}_i) \right)
 \ + \ \bm{P}_{trans},
\end{equation}
where use has been made of the relation $\bm{A} (\bm{r}_i) - \bm{A}_\perp (\bm{r}_i) = \bm{A}_\parallel (\bm{r}_i)$.  
The discussion so far is totally independent of the choice of gauge.
To make the following discussion as transparent as possible, we shall work
for a while in a particular gauge, i.e. the Coulomb gauge, and will come back
to more general case later. As was already explained, in the 
Coulomb gauge, we can set $\bm{A}_\parallel = 0$ without loss of
generality. The above expression for the total momentum $\bm{P}$ then
reduces to a very simple form given as 
\begin{equation}
 \bm{P} \ = \ \sum_i \,\bm{p}_i \ + \ \bm{P}_{trans}. 
\end{equation}
One observes that the total momentum of the charged particles and the photons
is given as a sum of the canonical momenta of charged particles and
the transverse momentum of the electromagnetic fields.

Next, let us consider a similar decomposition of the total angular momentum.
The total angular momentum of the system is a sum of the mechanical
angular momentum of charged particles and the angular momentum of photon fields as  
\begin{equation}
 \bm{J} \ = \ \sum_i \, m_i \,\bm{r}_i \times \dot{\bm{r}}_i \ + \ 
 \int \,d^3 r \,\,\bm{r} \times (\bm{E} \times \bm{B}).
\end{equation}
Similarly as before, the total angular momentum of the electromagnetic fields
can be decomposed into longitudinal and transverse parts as
\begin{equation}
 \int \,d^3 r \,\,\bm{r} \times (\bm{E} \times \bm{B}) \ = \ 
 \bm{J}_{long} \ + \ \bm{J}_{trans},
\end{equation}
with
\begin{eqnarray}
 \bm{J}_{long} &=& \int \,d^3 r \,\,\bm{r} \times 
 \left( \bm{E}_\parallel \times \bm{B}_\perp \right), \\
 \bm{J}_{trans} &=& \int \,d^3 r \,\,\bm{r} \times
 \left( \bm{E}_\perp \times \bm{B}_\perp \right) ,
\end{eqnarray}
which leads to the relation
\begin{equation}
 \bm{J} \ = \ 
 \sum_i \,m_i \,\bm{r}_i \times \dot{\bm{r}}_i \ \ + \ \ 
 \bm{J}_{long} \ + \ \bm{J}_{trans}.
\end{equation}
Again, by using the Gauss law, $\bm{J}_{long}$ can also be expressed as
\begin{equation}
 \bm{J}_{long} \ = \ \sum_i \,q_i \,\,\bm{r}_i \times \bm{A}_\perp (\bm{r}_i),
\end{equation}
so that we can write as
\begin{equation}
 \bm{J} \ = \ 
 \sum_i \,m_i \,\,\bm{r}_i \times \dot{\bm{r}}_i \ + \ \sum_i \,
 q_i \,\bm{r}_i \times \bm{A}_\perp (\bm{r}_i) \ + \ \bm{J}_{trans}. 
 \label{QED_JO}
\end{equation}
We recall that the quantity $q_i \,\bm{r}_i \times \bm{A}_\perp (\bm{r}_i)$
appearing in the above decomposition just corresponds to what-we-call the
{\it potential angular momentum} \cite{Wakamatsu10}.
In the present context, it represents the
angular momentum that associates with the longitudinal (electric) field
generated by the charged particle $i$.
Again, if one combines it with the mechanical angular momentum of the charged
particle $i$, the total angular momentum $\bm{P}$ of the system is
represented as
\begin{equation}
 \bm{J} \ = \ \sum_i \,\bm{r}_i \times \left(\,\bm{p}_i \ - \ 
 q_i \,\bm{A}_\parallel (\bm{r}_i) \,\right) \ + \ \bm{J}_{trans},
\end{equation}
in general gauges, and as
\begin{equation}
 \bm{J} \ = \ \sum \,\bm{r}_i \times \bm{p}_i \ + \ \bm{J}_{trans},
\end{equation}
in the Coulomb gauge.

Summarizing the above manipulations, we find (in the Coulomb gauge)
the following much simpler-looking expressions for the total momentum
and the total angular momentum of the interacting system of charged particles
and the photons.
\begin{eqnarray}
 \bm{P} &=& \sum_i \,\bm{p}_i \ + \ \bm{P}_{trans}, \\
 \bm{J} &=& \sum_i \,\bm{r}_i \times \bm{p}_i \ + \ 
 \bm{J}_{trans} .
\end{eqnarray}
At first sight, it appears to indicate physical superiority of canonical
momentum and the canonical angular momentum over the mechanical ones.
However, such a conclusion is premature, as is clear from the following
consideration of the energy of the system.
As already pointed out, the total Hamiltonian of the system is
given as a sum of three terms, i.e. the mechanical energies of the charged
particles, the Coulomb energies between them, and the energy of
transverse photons.
An important observation here is that, different from the cases of
total momentum and angular momentum, when the sum of the mechanical
energy and the Coulomb energy (the energy associate with $\bm{E}_\parallel$)
is expressed in terms of the canonical momentum, it does not reduce to a
simple form, because
\begin{equation}
 \frac{1}{2} \,m_i \,\dot{\bm{r}}_i^2 \ + \ V_{Coul} \ \neq \ 
 \sum_i \,\frac{\bm{p}_i^2}{2 \,m_i} .
\end{equation}
Instead, we have
\begin{eqnarray}
 H &=& \sum_i \,\frac{1}{2 \,m_i} \,
 \left(\,\bm{p}_i - q_i \,\bm{A}_\perp (\bm{r}_i) \right)^2 \ + \ 
 V_{Coul} \nonumber \\
 &=& \sum_i \frac{\bm{p}_i^2}{2 \,m_i} \ + \ V_{Coul} \nonumber \\
 &+& 
 \sum_i \,\frac{q_i}{2 \,m_i} \,\,[\,\bm{p}_i \cdot \bm{A}_\perp (\bm{r}_i)
 \ + \ 
 \bm{A}_\perp (\bm{r}_i) \cdot \bm{p}_i \,] \ + \ 
 \sum_i \,\frac{q_i^2}{2 \,m_i} \,
 \bm{A}_\perp (\bm{r}_i) \cdot \bm{A}_\perp (\bm{r}_i).
\end{eqnarray}
Crucially important to recognize here is the difference of the two
quantities,
\begin{equation}
 \sum_i \,\frac{1}{2} \,m_i \,\dot{\bm{r}_i}^2,
\end{equation}
and
\begin{equation}
 \sum_i \,\,\frac{\bm{p}_i^2}{2 \,m_i}.
\end{equation}
As already mentioned, the former quantity represents the mechanical
kinetic energy of charged particles, i.e. the kinetic energy
of particles which associate with their translational motion.
Usually, the latter quantity is also interpreted as the kinetic
energy of charged particles, which means that we are not distinguishing
these two quantities very clearly.
The reason is that we are too much accustomed with weakly coupled systems
of charged particles and photons.
To understand it, let us consider the problem of hydrogen atom.
Assuming, for simplicity, that the proton is infinitely heavy, it
reduces to a problem of one electron and photons,
described by the following Hamiltonian :
\begin{eqnarray}
 H &=& \frac{1}{2} \,m \,\dot{\bm{r}}^2 \ + \ V_{Coul} \ + \ H_{trans}
 \ = \ H_0 \ + \ H_{trans} \ + \ H_{int},
\end{eqnarray}
with
\begin{eqnarray}
 H_0 &=& \frac{\bm{p}^2}{2 \,m} \ + \ V_{Coul} (r), \\
 H_{trans} &=& \sum_{\bm{k}} \,\sum_{\lambda = 1,2} \,\hbar \,\omega_{\bm{k}}
 \,\,a^\dagger_{\bm{k},\lambda} \,a_{\bm{k},\lambda}, \\
 H_{int} &=& \frac{e}{2 \,m} \,[\,\bm{p} \cdot \bm{A}_\perp (\bm{r}) \ + \ 
 \bm{A}_\perp \cdot \bm{p} \,] \ + \ 
 \frac{e^2}{2 \,m} \,\bm{A}_\perp (\bm{r}) \cdot \bm{A}_\perp (\bm{r}) .
\end{eqnarray}
Here, $H_0$ is taken as an unperturbed Hamiltonian of the Hydrogen atom
with the Coulomb interaction between the electron and the proton,
$H_{trans}$ is the Hamiltonian of the transverse photons, and $H_{int}$
describes the interactions between the electron and the transverse photons.
A general form of eigen-states of the above Hamiltonian is expressed as
a direct product of the eigen-states of $H_0$ and those of $H_{trans}$
as $| \psi_n \rangle \otimes | \{ n_{\bm{k},\lambda} \} \rangle$,
where
\begin{equation}
 H_0 \,| \psi_n \rangle \ = \ E_n \,|\, \psi_n \rangle,
\end{equation}
while $| \{ n_{\bm{k},\lambda} \} \rangle$ is an abbreviation of the
following occupation-number representation of transverse photons :
\begin{equation}
 | \{ n_{\bm{k},\lambda} \} \rangle \ = \ \prod_\alpha \,
 | n_{\bm{k}_\alpha, \lambda_\alpha} \rangle .
\end{equation}
It is important to recognize that, in the ordinary description of hydrogen atom,
one does not include Fock components of transverse (real) photons.
(The formation of hydrogen atom is entirely due to the Coulomb
attraction between the proton and the electron, and the transverse photons
have little to do with it.)
Consequently, either of the total momentum or the total angular momentum
of the hydrogen atom is saturated by the electron alone, and the photons
carry none of them.
This also means that there is no practical difference between the mechanical
momentum
\begin{equation}
 \bm{P}_{mech} \ = \ m \,\dot{\bm{r}} \ = \ 
 \bm{p} - e \,\bm{A}_\perp,
\end{equation}
and the canonical momentum
\begin{equation}
 \bm{P}_{can} \ = \ \bm{p},
\end{equation}
since the expectation value of $\bm{A}_\perp$ in such restricted Fock space
is vanishing. Exactly the same can be said for the difference between the
mechanical angular momentum
\begin{equation}
 \bm{J}_{mech} \ = \ m \,\bm{r} \times \dot{\bm{r}} \ = \ 
 \bm{r} \times \left(\bm{p} - e \,\bm{A}_\perp \right),
\end{equation}
and the canonical angular momentum
\begin{equation}
 \bm{J}_{can} \ = \ \bm{r} \times \bm{p}.
\end{equation}
The difference between the mechanical kinetic energy
\begin{equation}
 \frac{1}{2} \,m \,\dot{\bm{r}}^2 
 \ = \ \frac{1}{2 \,m} \,
 \left( \bm{p} - e \,\bm{A}_\perp \right)^2 
\end{equation}
and the kinetic energy
\begin{equation}
 \frac{1}{2 \,m} \,\bm{p}^2
\end{equation}
is also ineffective in the static properties of the Hydrogen atom.
The fact is that the difference between these two quantities is
nothing but the interaction Hamiltonian, which is treated perturbatively,
thereby describing the processes of emissions and absorptions as well as
the scatterings of transverse photons by the hydrogen atom.

One must recognize that the situation is totally different in QCD.
Here, the nucleon is a strongly coupled gauge-system of
quarks and gluons. One certainly needs to include Fock components of
transverse gluons. Otherwise, the concept like the gluon distributions
in the nucleon would never be invoked. In such circumstances, 
the difference between the mechanical angular momentum and the
canonical angular momentum as well as the difference between the mechanical
momentum and the canonical momentum are generally nonzero and may
have sizable magnitude.

So far, we were mainly working in the Coulomb gauge, in order to
avoid unnecessary complexities for the above physical consideration.
Now we consider the problem of gauge-invariance more seriously.
As we shall see below, it provides us with a new and interesting insight
into the decomposition problem of the total momentum as well as the total
angular momentum of the interacting system of charged particles and photons.
(Since the argument goes in entirely the same manner for both of the total
momentum and the total angular momentum, we concentrate below on more
interesting angular momentum case.)
We have already shown that the total angular momentum $\bm{J}$ can be
decomposed into the following form in an arbitrary gauge :
\begin{equation}
 \bm{J} \ = \ \sum_i \,\bm{r}_i \times 
 (\,\bm{p}_i \ - \ q_i \,\bm{A}_\parallel (\bm{r}_i) \,) 
 \ + \ \bm{J}_{trans} .
\end{equation}
It is a well-known fact that the transverse part $\bm{J}_{trans}$ of photons
can further be decomposed into the orbital and spin parts as
\begin{equation}
 \bm{J}_{trans} \ = \ \int \,d^3 r \,\,E^l_\perp \,(\bm{r} \times \nabla) \,
 A^l_\perp \ + \ \int \,d^3 r \,\bm{E}_\perp \times \bm{A}_\perp.
\end{equation}
We emphasize that this decomposition is gauge-invariant, because $\bm{A}_\perp$
is gauge-invariant. (Naturally, $\bm{E}_\perp$ is gauge-invariant.)
Then, we are led to a decomposition as follows : 
\begin{equation}
 \bm{J} \ = \ \sum_i \,\bm{r}_i \times 
 (\bm{p}_i \ - \ q_i \,\bm{A}_\parallel (\bm{r}_i)) \ + \ 
 \int \,d^3 r \,\,E^l_\perp \,(\bm{r} \times \nabla) \,
 A^l_\perp \ + \ \int \,d^3 r \,\bm{E}_\perp \times \bm{A}_\perp.
 \label{QED_J}
\end{equation}
When going to quantum theory (in the coordinate representaion), the canonical
momentum is replaced by a differential operator as
\begin{equation}
 \bm{p}_i - q_i \,\bm{A}_\parallel (\bm{r}_i) \ \Rightarrow \ 
 \frac{1}{i} \,\left(\, \nabla_i \ - \ i \,q_i \,
 \bm{A}_\parallel (\bm{r}_i) \,\right).
\end{equation}
Notice that, with the identification $\bm{A}_\parallel = \bm{A}_{pure}$,
the r.h.s. is basically the pure-gauge derivative 
\begin{equation}
 \bm{D}_{i,pure} \ = \  
 \nabla_i \ - \ i \,q_i \,
 \bm{A}_{pure} (\bm{r}_i).
\end{equation}
introduced by Chen et al. \cite{Chen08},\cite{Chen09}.
Using it, Eq.(\ref{QED_J}) can now be written as
\begin{equation}
 \bm{J} \ = \ \bm{L}_p^\prime \ + \ \bm{L}_\gamma^\prime \ + \ 
 \bm{S}_\gamma^\prime ,
 \label{QED_J_decompII}
\end{equation}
with
\begin{eqnarray}
 \bm{L}_p^\prime &=& 
 \sum_i \,\bm{r}_i \times \frac{1}{i} \,\bm{D}_{i,pure}, \\
 \bm{L}_\gamma^\prime &=& \int \,d^3 r \,\,E^l_\perp \,
 (\bm{r} \times \nabla) \, A^l_\perp, \\
 \bm{S}_\gamma^\prime &=& \int \,d^3 r \,\,\bm{E}_\perp \times \bm{A}_\perp.
\end{eqnarray}
One may recognize now that this just corresponds to a gauge-invariant
decomposition of Chen et al. in the case of QED except that we are handling
 here the charged particles without intrinsic spin \cite{Chen08},\cite{Chen09}.
In fact, the gauge-invariance of the 1st term can readily be verified, by
using the gauge transformation property of the longitudinal component of
$\bm{A}_\parallel$
\begin{equation}
 \bm{A}_\parallel (\bm{r}_i) \ \rightarrow \ 
 \bm{A}_\parallel (\bm{r}_i) \ + \ \nabla \,\Lambda (\bm{r}_i),
\end{equation}
and the gauge transformation property of the quantum-mechanical wave
function of the charged particles given as
\begin{equation}
 \Psi (\bm{r}_1, \cdots, \bm{r}_N) \ \rightarrow \ 
 \left(\, \prod_i^N \,e^{i \,q_i \,\Lambda (\bm{r}_i)} \,\right) \,
 \Psi (\bm{r}_1, \cdots, \bm{r}_N) .
\end{equation}
As is obvious from our previous studies \cite{Wakamatsu10},\cite{Wakamatsu11A},
however, the above decomposition
(\ref{QED_J_decompII}) is not a unique possibility of gauge-invariant
decomposition of the
total angular momentum. 
To confirm it, we go back to Eq.(\ref{QED_JO}), which we now write as
\begin{eqnarray}
 \bm{J} &=& \sum_i \,m_i \,\bm{r}_i \times \dot{\bm{r}}_i \ + \ 
 \int \,d^3 r \,\,\bm{r} \times (\bm{E}_\parallel \times \bm{B}_\perp) 
 \nonumber \\
 &+& \int \,d^3 r \,\,E^l_\perp \,(\bm{r} \times \nabla) \,
 A^l_\perp \ + \ \int \,d^3 r \,\bm{E}_\perp \times \bm{A}_\perp 
\end{eqnarray}
Combining the piece $\int \,d^3 r \,\bm{r} \times (\bm{E}_\parallel
\times \bm{B}_\perp)$, which was previously written as
$\sum_i \,q_i \,\bm{r}_i \times \bm{A}_\perp (\bm{r}_i)$, with
the orbital part of $\bm{J}_{trans}$, we are led to another
decomposition : 
\begin{equation}
 \bm{J} \ = \ \bm{L}_p \ + \ \bm{L}_\gamma \ + \ \bm{S}_\gamma ,
\end{equation}
where
\begin{eqnarray}
 \bm{L}_p &=& \sum_i \,m_i \,\bm{r}_i \times \dot{\bm{r}}_i , \\
 \bm{L}_\gamma &=& \int \,d^3 r \,E^k_\perp \,
 (\bm{r} \times \nabla) \,A^k_\perp \ + \ 
 \int \,d^3 r \,\,\bm{r} \times (\bm{E}_\parallel \times \bm{B}_\perp), \\
 \bm{S}_\gamma &=& \int \,d^3 r \,\bm{E}_\perp \times \bm{B}_\perp.
\end{eqnarray}
Note that, using the relationship
$\bm{p}_i = m_i \,\dot{\bm{r}}_i + q_i \,\bm{A} (\bm{r}_i)$,
$\bm{L}_p$ can also be written as
\begin{eqnarray}
 \bm{L}_p &=& \sum_i \bm{r}_i \times (\,\bm{p}_i - q_i \,\bm{A} (\bm{r}_i) \,) \\
 &\rightarrow& \sum_i \bm{r}_i \times \frac{1}{i} \,
 \left(\,\nabla_i - i \,q_i \,\bm{A} (\bm{r}_i) \right) \ \equiv \
 \sum_i \,\bm{r}_i \times \frac{1}{i} \,\bm{D}_i . 
\end{eqnarray}
Obviously, this decomposition is also gauge-invariant. 
This gauge-invariant decomposition falls into the category of
decomposition (I), while the previous decomposition into that of
decomposition (II) according to the classification in \cite{Wakamatsu11A}.
As is clear by now, the difference between the two decompositions arises
from the treatment of the potential angular momentum term
$\sum_i \,q_i \,\bm{r}_i \times \bm{A}_\perp (\bm{r}_i)$, which is
solely gauge-invariant. In the decomposition (I), it
is included in the orbital angular momentum part of photons, while in the
decomposition (II), it is included in the orbital angular momentum
part of charged particles. 
As a consequence, what appears in the decomposition (I) is the mechanical
(or dynamical) angular momentum given as
\begin{equation}
 \bm{L}^{mech}_p \ \equiv \ \bm{L}_p \ = \ 
 \sum_i \,\bm{r}_i \times \frac{1}{i} \,\bm{D}_i \ \equiv \ 
 \sum_i \,\bm{r}_i \times \frac{1}{i} \,\left(\,
 \nabla_i - i \,q_i \,\bm{A} (\bm{r}_i) \right),
\end{equation}
containing full gauge-covariant derivative, while what appears in the
decomposition (II) is a generalized canonical angular momentum (with
gauge-invariance) given by
\begin{equation}
 \bm{L}^{``can"}_p \ \equiv \ \bm{L}_p^\prime \ = \ 
 \sum_i \,\bm{r}_i \times \frac{1}{i} \,
 \bm{D}_{i, pure} \ \equiv \ \sum_i \,\bm{r}_i \times \frac{1}{i} \,
 \left(\, \nabla_i - i \,q_i \,\bm{A}_\parallel (\bm{r}_i) \right), 
\end{equation}
which reduces to the ordinary canonical momentum in the Coulomb gauge,
in which $\bm{A}_\parallel (\bm{r}_i) = 0$. All of these are
anticipated facts from the analysis in our previous
papers \cite{Wakamatsu10},\cite{Wakamatsu11A}.
Here, we can say more. It is a wide-spread belief that, among the
two quantities, i.e. the canonical angular momentum and the
dynamical (or mechanical) angular momentum, what is closer to simple
physical image of orbital motion is the former because it appears that
the latter contains an {\it extra interaction term} between
the charged particles and the photons. (This prepossession is further
amplified by a simpler commutation relation of $\bm{L}_p^\prime$, which
is not possessed by $\bm{L}_p$ \cite{Sun10},\cite{Goldman11}.)
We now realize that the truth is just opposite.
In fact, we have shown that the {\it canonical} angular momentum is a
sum of the mechanical angular momentum and the longitudinal part
of the photon angular momentum as
\begin{eqnarray}
 \bm{L}_p^\prime &=& \bm{L}_p \ + \ \sum_i \,\bm{r}_i \times 
 q_i \,\bm{A}_\perp (\bm{r}_i) \\
 &=& \bm{L}_p \ + \ \int \,d^3 r \,\,\bm{r} \times
 (\bm{E}_\parallel \times \bm{B}_\perp) ,
\end{eqnarray}
where
\begin{equation}
 \bm{L}_p \ = \ \sum_i \,m_i \,\bm{r}_i \times \dot{\bm{r}}_i
 \ = \ \sum_i \,m_i \,\bm{r}_i \times \bm{v}_i.
 \label{mechanicalOAM}
\end{equation}
As is clear from the above expression (\ref{mechanicalOAM}) of $\bm{L}_p$,
it is the {\it mechanical} angular momentum $\bm{L}_p$ not the {\it canonical}
angular momentum $\bm{L}_p^\prime$ that
has a natural physical interpretation as orbital motion of particles.
It may really sound paradoxical, but what contains an extra interaction
term is rather the {\it canonical} angular momentum not the {\it mechanical}
angular momentum !

\section{Relation between the two inequivalent decompositions of the
nucleon momentum}

In one of the two papers \cite{Chen08},\cite{Chen09}, which brought
about a big argument on the
nucleon spin decomposition problem, Chen et al. suspect a
common wisdom of deep-inelastic-scattering (DIS) physics that
the gluons carry about half of the nucleon momentum in the asymptotic
limit. According to them, this large fraction is due to an unsuitable
definition of the gluon momentum in an interacting theory.
It was claimed that, if the quark and gluon momenta are defined in
a gauge-invariant and consistent way, the asymptotic limit of the
gluon momentum fraction would be only about one-fifth as compared
with the standardly-believed value of one-half. We shall inspect
below the validity of this astounding conclusion.

Their argument starts with the statement that the conventional gluon
momentum fraction is based on the following decomposition of the
total momentum operator in QCD : 
\begin{eqnarray}
 \bm{P}_{total} &=& \int \,d^3 x \,\psi^\dagger \,\frac{1}{i} \,
 \bm{D} \,\psi \ + \ \int \,d^3 x \,\bm{E} \times \bm{B} \nonumber \\
 &=& \hspace{10mm} \bm{P}_q \hspace{10mm} + \hspace{10mm} \bm{P}_G ,
\end{eqnarray}
where $\bm{D} = \nabla - i \,g \,\bm{A}$ is the standard covariant derivative.
The scale evolution of $\bm{P}_q$ and $\bm{P}_G$ is governed by
the following anomalous dimension matrix at the leading
order \cite{GW74},\cite{GP74} : 
\begin{equation}
 \gamma^P \ \equiv \ 
 \left( \begin{array}{cc}
 \gamma^{(2)}_{qq} & \gamma^{(2)}_{qG} \\
 \gamma^{(2)}_{Gq} & \gamma^{(2)}_{GG} \\
 \end{array} \right)
 \ = \ 
 \frac{\alpha_S}{8 \,\pi} \,
 \left( \begin{array}{cc}
 - \,\frac{8}{9} \,n_g & \frac{4}{3} \,n_f \\
 \frac{8}{9} \,n_g & - \,\frac{4}{3} \,n_f \\
 \end{array} \right) ,
\end{equation}
with $n_g$ and $n_f$ being the number of gluon fields and the number of
active quark flavors. This leads to the well-known asymptotic limit
for the gluon momentum fraction,
\begin{equation}
 \bm{P}_G \ = \ \frac{2 \,n_g}{2 \,n_g + 3 \,n_f} \,\bm{P}_{total}.
\end{equation}
Their objection to this common knowledge is based on another gauge-invariant
decomposition proposed by themselves : 
\begin{equation}
 \bm{P}_{total} \ = \ \bm{P}^\prime_q \ + \ \bm{P}^\prime_G,
\end{equation}
where
\begin{eqnarray}
 \bm{P}^\prime_q &=& \int \,d^3 x \,\,\psi^\dagger \,\frac{1}{i} \,
 \bm{D}_{pure} \,\psi , \\
 \bm{P}^\prime_G &=& \int \,d^3 x \,\,E^k \,
 \mbox{\boldmath ${\cal D}$}_{pure} \,A^k_{phys} ,
\end{eqnarray}
with
\begin{eqnarray}
 D^\mu_{pure} &\equiv& \partial^\mu \ - \ i \,g \,A^\mu_{pure}, \\
 {\cal D}^\mu_{pure} &\equiv& \partial^\mu \ - \ i \,g \,
 \left[\,A^\mu_{pure}, \ \cdot \ \,\right] .
\end{eqnarray}
Although the detail of the calculation was not shown, they concluded that
this decomposition leads to the following anomalous dimension
matrix \cite{Chen09}
\begin{equation}
 \gamma^{P^\prime} \ = \ 
 \frac{\alpha_S}{8 \,\pi} \,
 \left( \begin{array}{cc}
 - \,\frac{2}{9} \,n_g & \frac{4}{3} \,n_f \\
 \frac{2}{9} \,n_g & - \,\frac{4}{3} \,n_f \\
 \end{array} \right) ,
\end{equation}
thereby predicting a totally different asymptotic limit for the gluon momentum fraction,
\begin{equation}
 \bm{P}_G \ = \ \frac{n_g}{n_g + 6 \,n_f} \,\bm{P}_{total}.
\end{equation}
For the typical case of $n_f = 5$, this gives $\bm{P}^\prime_G \simeq
\frac{1}{5} \,\bm{P}_{total}$, as compared with the prediction of the
standard scenario
$\bm{P}_G \simeq \frac{1}{2} \,\bm{P}_{total}$. 

Apparently, to discuss the momentum sum rule of QCD and its evolution,
it is more convenient to handle the problem in a covariant way.
Along the same line as explained in our previous
paper \cite{Wakamatsu10},\cite{Wakamatsu11A},
which established the fact that there exist two physically inequivalent
decompositions of the QCD angular momentum tensor, we can show that
there are two different decompositions of the QCD energy-momentum tensor,
both of which are gauge-invariant. The decomposition (I) contains in it
the standard decomposition given in the paper \cite{JM90} : 
\begin{equation}
 T^{\mu \nu} \ = \ T^{\mu \nu}_q \ + \ T^{\mu \nu}_G , \label{T}
\end{equation}
with
\begin{eqnarray}
 T^{\mu \nu}_q &=& \frac{1}{2} \,\bar{\psi} \,\left(\, 
 \gamma^\mu \,i \,D^\nu \ + \ \gamma^\nu \,i \,D^\mu \,\right) \psi , 
 \label{Tq} \\
 T^{\mu \nu}_G &=& - \,\mbox{Tr} \,\left[\, F^{\mu \alpha} \,D^\nu \,A_\alpha
 \ + \ F^{\nu \alpha} \,D^\mu \,A_\alpha \,\right] \ + \ 
 \frac{1}{2} \,g^{\mu \nu} \,\mbox{Tr} \,[ F^2 ] .
 \label{TG}
\end{eqnarray}
Since the second term of $T^{\mu \nu}_G$ contributes only to the boost and
does not contribute to the momentum sum rule of the nucleon, we shall drop
it in the following argument. It can be shown that,
up to a surface term, the gluon part can further be decomposed into two
gauge-invariant pieces as
\begin{eqnarray}
 T^{\mu \nu}_G &=& - \,\mbox{Tr} \,\left[\, 
 F^{\mu \alpha} \,D^\nu_{pure} \,A_{\alpha,phys} \ + \ 
 F^{\nu \alpha} \,D^\mu_{pure} \,A_{\alpha,phys} \,\right] \nonumber \\
 &\,& - \,\mbox{Tr} \,\left[\, D_\alpha \,F^{\mu \alpha} \,A^\nu_{phys}
 \ + \ D_\alpha \,F^{\nu \alpha} \,A^\mu_{phys} \,\right]  \ \ + \ \  
 \mbox{surface term}  .
 \label{TG_decomp}
\end{eqnarray}
Here, the 2nd term of the above equation is a covariant generalization of
the potential momentum term as discussed in sect.III.
Under the imposed gauge transformation
property of the physical and pure \-gauge components of the gluon fields
given by
\begin{eqnarray}
 A^\mu_{phys} (x) &\rightarrow& U(x) \,A^\mu_{phys} (x) \,U^\dagger (x), 
 \label{Gtr_phys} \\
 A^\mu_{pure} (x) &\rightarrow& U(x) \,\left[\,
 A^\mu_{pure} (x) \ + \ \frac{i}{g} \,\partial^\mu \,\right] \,U^\dagger (x),
 \label{Gtr_pure}
\end{eqnarray}
supplemented with the pure-gauge condition for the pure-gauge part of
$A^\mu$
\begin{equation}
 F^{\mu \nu}_{pure} \ \equiv \ \partial^\mu \,A^\nu_{pure} \ - \ 
 \partial^\nu \,A^\mu_{pure} \ + \ i \,g \,[ A^\mu_{pure}, A^\nu_{pure} ]
 \ = \ 0, \label{pure_gauge_cond}
\end{equation}
it is easy to show that each term of (\ref{TG_decomp}) is separately
gauge-invariant. 
Eqs. (\ref{T})-(\ref{TG}) combined with (\ref{TG_decomp}) gives
our gauge-invariant decomposition
(I) of the QCD energy-momentum
tensor. Since the potential momentum term is still contained in the gluon
part in this decomposition, it is practically the same as the standard
decomposition.

On the other hand, if one combines the potential angular momentum term
with the quark part by making use of the QCD equation of motion
$\left( D_\alpha \,F^{\mu \nu} \right)^a = g \,
\bar{\psi} \,\gamma^\mu \,T^a \,\psi$, one is led to another gauge-invariant
decomposition (II) of QCD energy-momentum tensor given as follows : 
\begin{equation}
 T^{\mu \nu} \ = \ T^{\prime \mu \nu}_q \ + \ T^{\prime \mu \nu}_G ,
\end{equation}
where
\begin{eqnarray}
 T^{\prime \mu \nu}_q &=& \frac{1}{2} \,\bar{\psi} \,
 \left(\, \gamma^\mu \,i \,D^\nu_{pure} \ + \ 
 \gamma^\nu \,i \, D^\mu_{pure} \,\right) \,\psi , \\
 T^{\prime \mu \nu}_G &=& - \,\mbox{Tr} \,\left[\,
 F^{\mu \alpha} \,D^\nu_{pure} \,A_{\alpha,phys} \ + \ 
 F^{\nu \alpha} \,D^\mu_{pure} \,A_{\alpha,phys} \,\right] .
\end{eqnarray}
This decomposition is thought of as a covariant generalization of the
decomposition of Chen et al.

The question is now whether these two decompositions of QCD energy-momentum
tensor lead to different predictions for the quark and gluon momentum
fractions and their evolution.
As emphasized in \cite{Wakamatsu11A}, a remarkable feature of our gauge-invariant
decompositions (I) and (II) is that we have not yet fixed gauge explicitly.
This means that we can choose any gauge as long as the choice is consistent
with the above-mentioned general conditions 
(\ref{Gtr_phys})-(\ref{pure_gauge_cond}). Particularly useful is the
fact that we can take the light-cone gauge as well \cite{Wakamatsu11A},
which is the most convenient gauge for discussing DIS observables.

We can then follow the argument given by Jaffe \cite{Jaffe01}.
The simplest way of
obtaining the momentum sum rule is to evaluate the nucleon matrix
element of the $(++)$-component of the energy-momentum tensor.
The momentum sum rule then follows from the normalization condition : 
\begin{equation}
 \frac{\langle P \,|\, T^{++} \,|\, P \rangle}{2 \,(P^+)^2} \ = \ 1.
 \label{Tpp_norm}
\end{equation}
As emphasized by Jaffe, $T^{++}$ simplifies dramatically in $A^+ = 0$ gauge,
because of the simplification of $D^+$ and $F^{+ \alpha}$,
\begin{eqnarray}
 D^+ &=& \partial^+ \ - \ i \,g \,A^+ \ \rightarrow \ \partial^+ , \\
 F^{+ \alpha} &=& \partial^+ \,A^\alpha \ - \ \partial^\alpha \,A^+ \ + \ 
 g \,[ A^+, A^\alpha ] \ \rightarrow \ \partial^+ \,A^\alpha .
\end{eqnarray}
As a consequence, $T^{++}$ reduces to a marvelously simple form as
\begin{eqnarray}
 T^{++} &=& T^{++}_q \ + \ T^{++}_G \nonumber \\
 &\rightarrow& \psi^\dagger_+ \,i \,\partial^+ \,\psi_+ \ + \ 
 2 \,\mbox{Tr} \,(\partial^+ \,\bm{A}_\perp)^2 ,
\end{eqnarray}
where $\psi_+$ is the standard $(+)$-component of $\psi$ defined by
$\psi_+ \equiv \ P_+ \,\psi$ and $P_+ = \frac{1}{2} \,\gamma^- \,\gamma^+$
with $\gamma^\pm = \frac{1}{\sqrt{2}} \,(\gamma^0 \pm \gamma^3)$. 
The two terms here give the contributions of quarks and gluons,
respectively, to $P^+$. Each term can be related to the 2nd moment of the
positive definite parton momentum distribution : 
\begin{eqnarray}
 \psi^\dagger_+ \,i \,\partial^+ \,\psi_+ \ \rightarrow \ 
 \int \,dx x \,q(x), \ \ \ \ 
 (\partial^+ \,\bm{A}^a_\perp)^2 \ \rightarrow \ 
 \int \,dx x \,g(x) .
\end{eqnarray}
The normarization condition (\ref{Tpp_norm}) then gives the well-known
momentum sum rule of QCD,
\begin{equation}
 1 \ = \ \int \,dx \,x \,[\,q(x, Q^2) \ + \ g(x, Q^2) \,] .
\end{equation}
This is a familiar story about the standard decomposition of the QCD
energy-momentum tensor.

A question is what would change if one adopts the decomposition (II), which
is thought to contain in it the decomposition of Chen et al.
To answer this question, we first recall the following relation
between the quark part of $T^{++}$ in the two
decompositions : 
\begin{equation}
 T^{++}_q \ - \ T^{\prime ++}_q \ = \ g \,\bar{\psi} \,\gamma^+ \,A^+_{phys} \,\psi .
 \label{Tpp_relation}
\end{equation}
We emphasize that the difference is nothing but a special component of
generalized potential momentum tensor.
Remember now the fact that, different from Chen et al.'s treatment,
we have a freedom to choose even the
light-cone gauge. Since $A^+ = A^+_{phys} = A^+_{pure} = 0$ in this gauge,
the difference between $T^{++}$ and $T^{\prime ++}$ simply vanishes.
We must therefore conclude that the two decompositions (I) and (II)
give exactly the same answer, as far as the
longitudinal momentum sum rule is concerned.
This fact has been verified in a particular gauge, i.e. in the light-cone gauge.
Note, however, that both of our decompositions (I) and (II) are manifestly
gauge-invariant. It is therefore a logical consequence of gauge-invariance
that the statement must hold in arbitrary gauges. 
(Naturally, it is of vital
importance to confirm the validity of this statement through explicit
calculations in other gauges than the light-cone gauge.)

Still, one might worry about the claim by Chen et al. that the two decompositions
of the nucleon momentum lead to totally different evolution equations for
the momentum fractions of quarks and gluons in the nucleon \cite{Chen09}.  
Let us next try to clarify this point.
Before discussing the evolution equation corresponding to the decomposition
(II), we think it useful to recall some basic knowledge on the
evolution matrix for the quark and gluon momentum fractions corresponding to
the standard decomposition (I).
Though somewhat trivial to remark, since the quark and the gluon parts of
this standard decomposition is separately gauge-invariant, the evolution
matrix should be independent of gauge choice. First we concentrate on
the quark part of $T^{++}$, which consists of two parts in general gauge as
\begin{equation}
 T^{++} \ = \ V_A \ + \ V_B,
\end{equation}
with
\begin{eqnarray}
 V_A &=& \bar{\psi} \,\gamma^+ \,i \,\partial^+ \,\psi , \\
 V_B &=& g \,\bar{\psi} \,\gamma^+ \,A^+ \,\psi .
\end{eqnarray}

\vspace{5mm}
\begin{figure}[ht]
\begin{center}
\includegraphics[width=13.5cm]{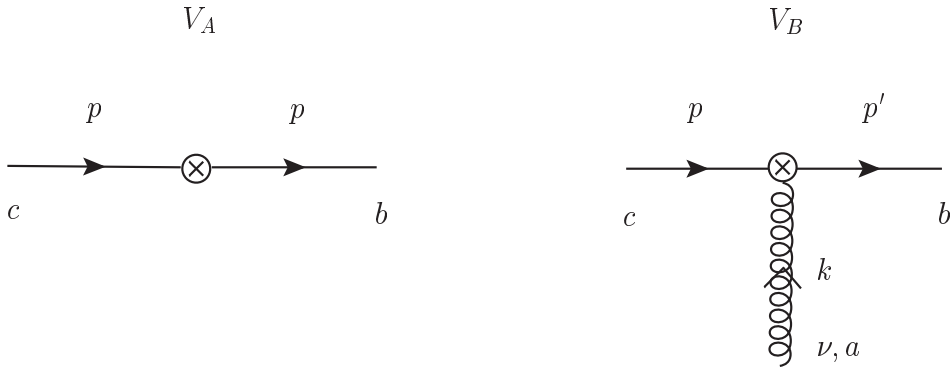}
\caption{Momentum space vertices for the quark part of $T^{++}$.}
\label{Fig:Vertex_VA_VB}
\end{center} 
\end{figure}

The momentum space vertices corresponding to these operators are
expressed by the following formulas supplemented  with the diagram
shown in Fig.\ref{Fig:Vertex_VA_VB} : 
\begin{eqnarray}
 V_A &=& \delta_{b c} \,\gamma^+ \,p^+ , \\
 V_B &=& g \,(T^a)_{b c} \,\gamma^+ \,g^{+ \nu} .
\end{eqnarray}
Note that $V_B \neq 0$ in general gauges, although $V_B = 0$ in the
light-cone gauge.

\vspace{3mm}
\begin{figure}[ht]
\begin{center}
\includegraphics[width=13.5cm]{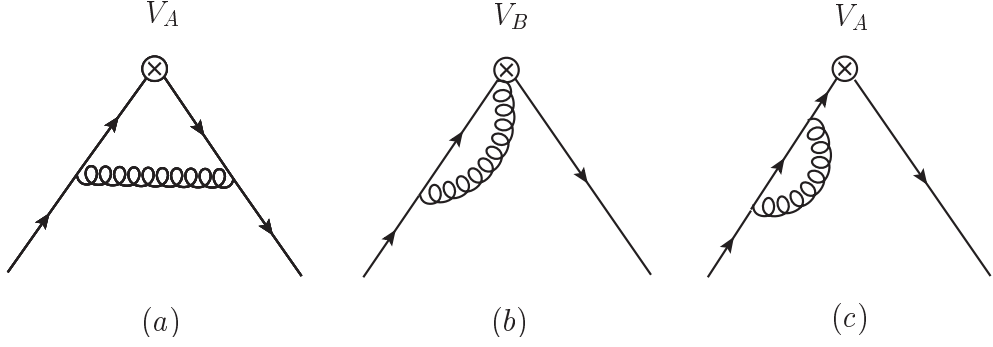}
\caption{One-loop diagrams contributing to the anomalous dimension of the
quark part of the energy momentum tensor. The diagram $(c)$
corresponds to the quark field-strength renormalization. Graphs that
are not symmetric with respect to the vertical lines through the operator
vertex have to be counted twice.}
\label{Fig:Emomentum}
\end{center} 
\end{figure}

Shown in Fig.\ref{Fig:Emomentum} are the Feynman diagrams, which contribute
to the anomalous
dimension $\gamma^{(2)}_{qq}$ in general covariant gauge. 
The answer in the Feynman gauge is well known. It is given by
\begin{equation}
 \gamma^{(2)}_{qq} \ = \ \frac{\alpha_S}{2 \,\pi} \,
 \left\{\,\frac{1}{6} \,C_F \ - \ C_F \ - \ \frac{1}{2} \,C_F \,\right\}
 \ = \ \frac{\alpha_S}{2 \,\pi} \,\left( - \,\frac{4}{3} \,C_F \right) ,
\end{equation}
with $C_F = \frac{4}{3}$. Here, the three terms in the middle of the above
equation respectively correspond to the contributions from the graphs
$(a), (b)$ and $(c)$ of Fig.\ref{Fig:Emomentum}.
On the other hand, in the light-cone gauge,
there is no contribution from the
graph $(b)$, and the answer is given as
\begin{equation}
 \gamma^{(2)}_{qq} \ = \ \frac{\alpha_S}{2 \,\pi} \,\left\{\,
 - \,\frac{17}{6} \,C_F \ + \ 0 \ + \ \frac{3}{2} \,C_F \,\right\}
 \ = \ \frac{\alpha_S}{2 \,\pi} \,\left( - \,\frac{4}{3} \,C_F \right) .
\end{equation}
Although individual term contributes differently,
the final answer is just the same as that of Feynman gauge.

Now that we have convinced that the anomalous dimension $\gamma^{(2)}_{qq}$
corresponding to the standard decomposition of the energy-momentum
tensor is independent of the choice of gauge, our next task is to
obtain the anomalous dimension $\gamma^{\prime (2)}_{qq}$
corresponding to another gauge-invariant decomposition (II) of the
QCD energy momentum tensor. For this purpose, we recall again the fact
that the quark parts of $T^{++}$ in the two decomposition are connected
through the relation (\ref{Tpp_relation}).
As pointed out before, the r.h.s. of (\ref{Tpp_relation})
vanishes in the light-cone-gauge.
This already indicates that the anomalous dimensions
corresponding to the two decompositions are the same, i.e.
$\gamma^{(2)}_{qq} = \gamma^{\prime (2)}_{qq}$. Let us verify this
statement more explicitly by showing that the vertex
$V_C \equiv g \,\bar{\psi} \,\gamma^+ \,A^+_{phys} \,\psi$ does not
contribute to the corresponding anomalous dimension even in other gauges
than the light-cone gauge. A key factor here is the fact that the
gluon field contained in the vertex $V_C$ is its physical part
$A^+_{phys}$. By taking care of this fact, we recall somewhat
nonstandard Feynman rule proposed in \cite{Wakamatsu11B}.
According to this rule, the
momentum representation of the vertex $V_C$ is given as
\begin{equation}
 V_C \ = \ g \,(T^a)_{b c} \,\gamma^+ \,g^{+ \nu} \,P^\nu_T ,
\end{equation}
which is delicately different from the vertex $V_B$ in that it contains a kind
of projection operator $P^\nu_T$. This projection operator $P^\nu_T$
with the Lorentz index $\nu$ reminds us of the fact that we must use the
modified gluon propagator
\begin{equation}
 \tilde{D}^{\mu \nu}_{a b} (k) \ = \ \frac{i \,\delta_{ab}}{k^2 + i \,\varepsilon}
 \,T^{\mu \nu},
\end{equation}
with
\begin{equation}
 T^{\mu \nu} \ = \ \sum_{\lambda = 1}^2 \,\varepsilon^\mu (k, \lambda) \,
 \varepsilon^{\nu *} (k, \lambda) ,
\end{equation}
whenever it is obtained with the contraction with the vertex $V_C$
containing the Lorentz index $\nu$. The Feynman diagram, which may potentially
contribute to the anomalous dimension in question, is given by the same graph
as the graph (b) of Fig.2 except that the vertex $V_B$ is replaced by $V_C$.   
An explicit calculation given in Appendix A shows that the contribution of
this diagram vanishes. 
(Note that this conversely means that the contribution to $\gamma^{(2)}_{qq}$
from the graph $(b)$ in the Feynman gauge comes totally from the vertex
$g \,\bar{\psi} \,\gamma^+ \,A^+_{pure} \,\psi$.)

\begin{figure}[ht]
\begin{center}
\includegraphics[width=5.5cm]{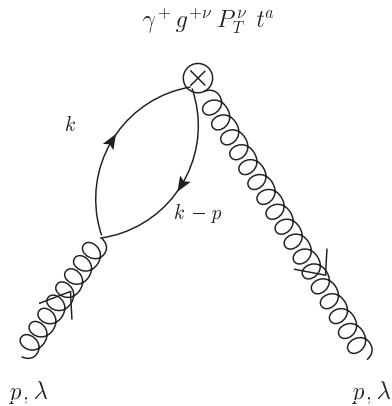}
\caption{The Feynman graph, which may potentially contribute to
the anomalous dimension $\gamma^{(2)}_{qG}$.}
\label{Fig:Mom_Vertex_qG}
\end{center} 
\end{figure}

Although slightly more trivial, we have also checked in Appendix A that
the potential momentum term does not contribute to the anomalous
dimension $\gamma^{(2)}_{qG}$. (The relevant diagram appearing in this proof
is illustrated in Fig.\ref{Fig:Mom_Vertex_qG}.) In this way, we now confirm
that
\begin{equation}
 \gamma^{(2)}_{qq} \ = \ \gamma^{\prime (2)}_{qq}, \ \ \ \ 
 \gamma^{(2)}_{qG} \ = \ \gamma^{\prime (2)}_{qG} .
\end{equation}

\vspace{3mm}

As is well-known, because of the conservation of total momentum, the $2 \times 2$
evolution matrix of the quark and gluon momenta (in whatever decomposition)
has only two independent elements such that
\begin{equation}
 \gamma^{(2)}_{Gq} \ = \ - \,\gamma^{(2)}_{qq}, \ \ \ \ 
 \gamma^{(2)}_{GG} \ = \ - \,\gamma^{(2)}_{qG} .
\end{equation}
We therefore conclude that the anomalus dimension matrix corresponding to
the two decompositions (I) and (II) of the QCD energy-momentum tensor
are exactly the same, i.e.
\begin{equation}
 \left( \begin{array}{cc}
 \gamma^{(2)}_{qq} & \gamma^{(2)}_{qG} \\
 \gamma^{(2)}_{Gq} & \gamma^{(2)}_{GG} \\
 \end{array} \right) \ = \ 
 \left( \begin{array}{cc}
 \gamma^{\prime (2)}_{qq} & \gamma^{\prime (2)}_{qG} \\
 \gamma^{\prime (2)}_{Gq} & \gamma^{\prime (2)}_{GG} \\
 \end{array} \right) .
\end{equation}
This contradicts the conclusion of Chen et al. given in \cite{Chen09}.
According to our analysis above, the gluons {\it do} carry about one-half of
the total nucleon momentum in the asymptotic limit.

The readers might suspect that the conclusion above contradicts our previous
statement that the Chen et al's decomposition is contained in our more
general decomposition (II), so that the physical predictions should be the
same. Note however that no one has yet checked the validity of their calculation
based on the Coulomb gauge.
Possible reasons of discrepancy might therefore be the following. One possibility
is that they have made a mistake in their Coulomb gauge calculation of
the anomalous dimension matrix. In fact,
the treatment of the gluon propagator in the Coulomb gauge is known to be
a fairly delicate issue because of the so-called energy-divergence in loop
integrations \cite{Leibbrandt94}.
Another possibility is that what they have calculated do not precisely
correspond to the evolution matrix of the longitudinal momentum
fractions of quarks and gluons appearing in deep-inelastic-scattering
physics.

\section{Relation between the two inequivalent decompositions of the
nucleon spin}

In the previous section, we have shown that there certainly exist two
generally inequivalent decompositions of the nucleon total momentum,
analogous to the QED problem. Nonetheless, as long as the longitudinal
momentum sum rules of the nucleon is concerned, the two decompositions
turn out to give completely the same answer for the quark and gluon
momentum fractions including their scale evolution.

Now, we turn to more interesting problem of nucleon spin decomposition.
We first recall the fact that there exist two different decompositions
also for the nucleon spin, both of which are gauge-invariant.
The QCD angular momentum tensor in the decomposition (I) is given by
Eq.(\ref{decompositionI}), while that in the decomposition (II) is given
by Eq.(\ref{decompositionII}).
The theoretical basis for obtaining the nucleon spin sum rule is given by
the equation \cite{Ji97PRD}
\begin{equation}
 \langle P s \,|\, W^\mu \,s_\mu \,| \,P s \rangle \,/ \,
 \langle P s \,|\, P s \rangle \ = \ \frac{1}{2},
\end{equation}
where $s_\mu$ is the covariant spin vector of the nucleon, while 
\begin{equation}
 W^\mu \ = \ - \,\epsilon^{\mu \nu \alpha \beta} \,J_{\alpha \beta} \,
 P_\gamma \,/\,(\,2 \,\sqrt{P^2}\,) ,
\end{equation}
with
\begin{equation}
 J^{\alpha \beta} \ = \ \int \,d^3 x \,\,M^{0 \alpha \beta} ,
\end{equation}
is the so-called Pauli-Lubansky vector \cite{Lubanski42}.
Assuming that the nucleon is
moving in the $z$ direction with momentum $P^\mu$ and helicity $+ \,1/2$,
it holds that
\begin{equation}
 J^{12} \,|\,P + \rangle \ = \ \frac{1}{2} \,|\,P + \rangle .
\end{equation}
Thus we are led to the relation
\begin{equation}
 \frac{1}{2} \ = \ \langle P + \,|\, J^{12} \,|\,P + \rangle \,/\,
 \langle P + \,|\, P + \rangle ,
\end{equation}
which provides us with a basis for obtaining longitudinal spin sum rule
of the nucleon. Depending on the two decompositions of $M^{\mu \nu \lambda}$,
this gives the following sum rules.
The decomposition (I) gives
\begin{equation}
 \frac{1}{2} \ = \ \left(\, \frac{1}{2} \,\Delta \Sigma \ + \ L_q \,\right)
 \ + \ \left(\, \Delta G \ + \ L_G \,\right) \ = \ J_q \ + \ J_G ,  
\end{equation}
where
\begin{eqnarray}
 \Delta \Sigma 
 &=& \langle P + \,|\, \int \,d^3 x \,\,
 \psi^\dagger \,\gamma^0 \,\gamma^3 \,\gamma_5 \,\psi \,|\,P + \rangle , \\
 L_q 
 &=& \langle P + \,|\, \int \,d^3 x \,\,
 \psi^\dagger \,(x^1 \,D^2 \ - \ x^2 \,D^1) \,\psi \,|\, P + \rangle, \\
 \Delta G 
 &=& \langle P + \,|\, \int \,d^3 x \,\,
 (E^1 \,A^2_{phys} \ - \ E^2 \,A^1_{phys} ) \,|\,P + \rangle , \\
 L_G 
 &=& \langle P + \,|\, \int \,d^3 x \,\,
 2 \,\,\mbox{Tr} \,\left\{\,E^k \,(x^2 \,{\cal D}^1_{pure} \ - \ 
 x^1 \,{\cal D}^2_{pure} ) \,A^k_{phys} \,\right\} \,|\,P + \rangle \nonumber \\
 &+& \langle P + \,|\, \int \,d^3 x \,\,2 \,\,\mbox{Tr} \,
 \left\{\,(\bm{D} \cdot \bm{E}) \,(x^2 \,A^2_{phys} \ - \ x^2 \,A^1_{phys})
 \,\right\} \,|\, P + \rangle ,
\end{eqnarray}
where we have neglected the normalization of the state, for simplicity.

On the other hand, the decomposition (II) leads to
\begin{equation}
 \frac{1}{2} \ = \ \left(\, \frac{1}{2} \,\Delta \Sigma^\prime 
 \ + \ L_q^\prime \,\right)
 \ + \ \left(\, \Delta G^\prime \ + \ L_G^\prime \,\right) 
 \ = \ J_q^\prime \ + \ J_G^\prime ,  
\end{equation}
where
\begin{eqnarray}
 \Delta \Sigma^\prime &=& \Delta \Sigma , \\
 L_q^\prime 
 &=& \langle P + \,|\, \int \,d^3 x \,\,
 \psi^\dagger \,(x^1 \,D^2_{pure} \ - \ x^2 \,D^1_{pure}) \,\psi \,|\, P + \rangle, \\
 \Delta G^\prime &=& \Delta G, \\
 L_G^\prime 
 &=& \langle P + \,|\, \int \,d^3 x \,\,
 2 \,\,\mbox{Tr} \,\left\{\,E^k \,(x^2 \,{\cal D}^1_{pure} \ - \ 
 x^1 \,{\cal D}^2_{pure} ) \,A^k_{phys} \,\right\} \,|\,P + \rangle.
\end{eqnarray}
The difference between the two decompositions resides in the orbital parts.
Note that $L_q$ and $L_q^\prime$ respectively correspond to the nucleon
matrix elements of mechanical and generalized canonical OAM operators.
What characterizes the difference of these two quantities is the
forward matrix element of the potential angular momentum given by
\begin{eqnarray}
 L_q \ - \ L_q^\prime &=& - \,(L_G \ - \ L_G^\prime ) \nonumber \\
 &=& \langle P + \,|\, \int \,d^3 x \,\,g \,\psi^\dagger \,
 (x^1 \,A^2_{phys} \ - \ x^2 \,A^1_{phys} ) \,\psi \,|\, P + \rangle .
 \label{Lq_relation}
\end{eqnarray}
It is important to recognize the fact that $A^1_{phys}$ and $A^2_{phys}$
are physical (transverse) components of gluons, which cannot be transformed
away even in the light-cone gauge. (This is totally different from the
case of the nucleon momentum decomposition, where the transverse components
do not appear in the difference between $T^{++}_q$ and $T^{\prime ++}_q$.)
Since the quantum state vector of the nucleon as a strongly-coupled gauge
system of quarks and gluons definitely contains Fock components of
transverse gluons, we conclude that the difference between $L_q$ and $L_q^\prime$
is generally non-zero. An explicit calculation by Burkardt and BC based on
simple models appears to confirm it \cite{BBC09}.

Nonetheless, one mysterious observation still remains to be clarified.
The problem concerns the scale dependence of quark and gluon OAMs.
Accepting that there are two different OAMs of both of quarks and gluons,
i.e. ($L_q, L_G)$ and $(L_q^\prime, L_G^\prime)$, one might
naturally expect different evolution equations for these two kinds of OAMs.
Somewhat embarrassingly, the past studies indicate that the evolution equations
of $L_q$ and $L_G$ are nothing different from those of $L_q^\prime$ and
$L_G^\prime$ \cite{JTH96}\nocite{HJL98}-\cite{Teryaev98}.
This can be confirmed as follows.
First, the scale dependence of $\Delta \Sigma$ and $\Delta G$ at the leading
order is widely known \cite{AP77},\cite{Sasaki75},\cite{AR76} and given as
\begin{equation}
 \frac{d}{d t} \,\left( \begin{array}{c}
 \Delta \Sigma \\
 \Delta G \\
 \end{array} \right) \ = \ 
 \frac{\alpha_S (t)}{2 \,\pi} \,
 \left( \begin{array}{cc}
 0 & 0 \\
 \frac{3}{2} \,C_F & \frac{\beta_0}{2} \\
 \end{array} \right) \,
 \left( \begin{array}{c}
 \Delta \Sigma \\
 \Delta G \\
 \end{array} \right) ,
\end{equation}
where $t = \ln Q^2 \,/\, \Lambda^2_{QCD}$, $C_F = 4/3$, and
$\beta_0 = 11 - \frac{2}{3} \,n_f$.
On the other hand, the leading-log evolution equation of quark and gluon
OAMs $L_q^\prime$ and $L_G^\prime$ was first derived by Ji, Tang, and
Hoodbhoy \cite{JTH96}. It is given by
\begin{eqnarray}
 \frac{d}{d t} \,\left( \begin{array}{c}
 L_q^\prime \\
 L_G^\prime \\
 \end{array} \right) 
 &=& \frac{\alpha_S (t)}{2 \,\pi} \,
 \left( \begin{array}{cc}
 - \,\frac{4}{3} \,C_F & \frac{n_f}{3} \\
 \frac{4}{3} \,C_F & - \,\frac{n_f}{3} \\
 \end{array} \right) \,
 \left( \begin{array}{c}
 L_q^\prime \\
 L_G^\prime \\
 \end{array} \right) \nonumber \\
 &+& \frac{\alpha_S (t)}{2 \,\pi} \,
 \left( \begin{array}{cc}
 - \,\frac{2}{3} \,C_F & \frac{n_f}{3} \\
 - \frac{5}{6} \,C_F & - \,\frac{11}{2} \\
 \end{array} \right) \,
 \left( \begin{array}{c}
 \Delta \Sigma \\
 \Delta G \\
 \end{array} \right) .
\end{eqnarray}
To be more precise, their derivation is based on a gauge-noninvariant
definition of $L^\prime_q$ and $L^\prime_G$ appearing in the Jaffe-Manohar
decomposition. Luckily, their calculation was done in the light-cone gauge.
This ended up with the result that the derived evolution equation coincides
with the answer obtained from the gauge-invariant definition of
$L^\prime_q$ and $L^\prime_G$ appearing in our decompositin (II).
(Remember the similar situation which we encounter in the study of
evolution equation of $\Delta G$ \cite{Wakamatsu11B}.
The point is that the Jaffe-Manohar decomposition is now taken as a
gauge-fixed form of our more general decomposition with manifest
gauge-invariance.)

Using the above evolution equations for $(\Delta \Sigma, \Delta G)$ and
$(L^\prime_q, L^\prime_G)$, one can easily write down the evolution
equation of the quark and gluon total angular momentum in the decomposition (II),
which are defined by $J_q^\prime \equiv L_q^\prime + \frac{1}{2} \,\Delta \Sigma$
and $J_G^\prime \equiv L_G^\prime + \Delta G$.
One finds that
\begin{equation}
 \frac{d}{d t} \,\left( \begin{array}{c}
 J_q^\prime \\
 J_G^\prime \\
 \end{array} \right) \ = \ \frac{\alpha_S (t)}{2 \,\pi} \,
 \left( \begin{array}{cc}
 - \,\frac{4}{3} \,C_F & \frac{n_f}{3} \\
 \frac{4}{3} \,C_F & - \,\frac{n_f}{3} \\
 \end{array} \right) \,
 \left( \begin{array}{c}
 J_q^\prime \\
 J_G^\prime \\
 \end{array} \right) .
\end{equation}
As noticed by several authors \cite{HJL98},\cite{Teryaev98},
the evolution matrix appearing here is just the
same as that of the momentum fractions of quarks and gluons.
On the other hand, Ji showed that the scale evolution of the total angular
momenta of quarks and gluons appearing in the decomposition (I) is
controlled by the same evolution matrix as that of the quark and gluon
momentum fractions as 
\begin{equation}
 \frac{d}{d t} \,\left( \begin{array}{c}
 J_q \\
 J_G \\
 \end{array} \right) \ = \ \frac{\alpha_S (t)}{2 \,\pi} \,
 \left( \begin{array}{cc}
 - \,\frac{4}{3} \,C_F & \frac{n_f}{3} \\
 \frac{4}{3} \,C_F & - \,\frac{n_f}{3} \\
 \end{array} \right) \,
 \left( \begin{array}{c}
 J_q \\
 J_G \\
 \end{array} \right) .
\end{equation}
The reason is that the quark and gluon angular momenta
$J_q$ and $J_G$ in the decomposition (I)
are defined by the QCD angular momentum tensor $M^{\alpha \mu \nu}$,
which is related to the energy-momentum tensor
$T^{\mu \nu}$ through the relation
\begin{equation}
 M^{\alpha \mu \nu} \ = \ T^{\alpha \nu} \,x^\mu \ - \ T^{\alpha \mu} \,x^\nu ,
\end{equation}
with
\begin{equation}
 T^{\mu \nu} \ = \ T^{\mu \nu}_q \ + \ T^{\mu \nu}_G .
\end{equation}
According to Ji, forming spatial moment of $T^{\mu \nu}_q$ and
$T^{\mu \nu}_G$ does not change the short-distance singularity of the operators.
It then follows that
$(J_q, J_G)$ and $(\langle x \rangle_q, \langle x \rangle_G)$ obey the
same evolution equation.
At any rate, one now realizes that $(J_q, J_G)$ and $(J_q^\prime, J_G^\prime)$
obey the same evolution equation at least at the one-loop order.
Since $\Delta \Sigma$ and $\Delta G$ are common in the two decompositions,
this also means that $(L_q, L_G)$ and $(L_q^\prime, L_G^\prime)$ obey the
same evolution equation.

\vspace{3mm}
\begin{figure}[ht]
\begin{center}
\includegraphics[width=5.5cm]{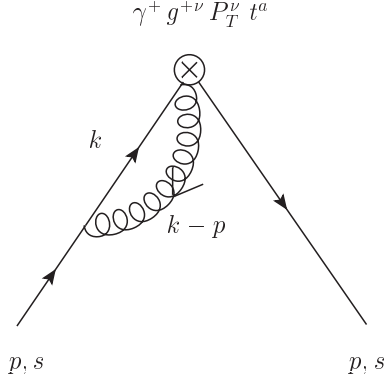}
\caption{The Feynman graph, which may potentially contribute to
the evolution matrix for the quark orbital angular momentum.}
\label{Fig:Vertex_PotenOAM}
\end{center} 
\end{figure}

How can we understand this somewhat puzzling observation ?
The answer is basically given in the paper by Ji \cite{Ji97PRL}.
He claims that the above observation can be understood, since the
interaction-dependent term, $g \,\int \,d^3 x \,\psi^\dagger \,
\bm{x} \times \bm{A} \,\psi$, which characterizes the difference between the
dynamical and canonical angular momenta of quarks, shall not affect the
leading-log evolution in the light-cone gauge.
Unfortunately, an explicit proof is not given there.
Furthermore, the statement holds only in the light-cone gauge, because
it is based on gauge-noninvariant expression
$g \,\int \,d^3 x \,\psi^\dagger \,\bm{x} \times \bm{A} \,\psi$ of
the interaction-dependent part.
To refine the statement and also to make the role of gauge-invariance more
manifest, we recall the fact that the difference of $L_q$ and $L_q^\prime$ is
given by the nucleon matrix element of potential angular momentum 
(see (\ref{Lq_relation})), which is a manifestly gauge-invariant quantity. 
It is therefore possible to extend the validity of Ji's statement
by showing that the (gauge-invariant) potential angular momentum term
does not contribute to the evolution matrix also in other gauges than
the light-cone gauge. The proof is given in Appendix B. (The relevant
Feynman diagram is shown in Fig.\ref{Fig:Vertex_PotenOAM}.)
This clarifies the reason why $(L_q, L_G)$ and $(L^\prime_q, L^\prime_G)$,
appearing in the two generally different decompositions of the nucleon spin,
obey the same evolution equation.

To avoid misunderstanding, we want to reemphasize the following fact.
In the case of longitudinal momentum sum rule discussed in the previous
section, we showed that the two decompositions of the QCD energy-momentum
tensor gives the same evolution equation for the momentum fractions of
quarks and gluons. In this case, the numerical values of the
quark and gluon momentum fractions in the two decompositions are also
the same at an arbitrary energy scale, because the transverse
components of the gluon fields never contribute to the longitudinal
momentum sum rule, as can be convinced from the
expression (\ref{Tpp_relation}).
This is not the case for
the longitudinal spin sum rule of the nucleon, however. Although the
quark and gluon OAMs appearing in the two decompositions (I) and (II)
are shown to obey the same evolution equation, there is no reason that
their numerical values at an arbitrary energy scale also coincide.
In fact, they are generally different, because the transverse (real) gluon fields
{\it do} contribute to the difference between the two definitions of quark
and gluon OAMs in the nucleon. (Remember the relation (\ref{Lq_relation}).)
As emphasized in \cite{Waka10}, a clear recognition of this fact is
especially important, if one tries to compare the predictions of low energy
effective models on the nucleon spin contents with those of lattice
QCD \cite{Waka10}\nocite{MT88}\nocite{Thomas08}\nocite{WT05}\nocite{WN06}
\nocite{WN08}\nocite{LHPC08}-\cite{LHPC10}.

\section{Summary and conclusion}

In summary, we first briefly review the current status of the nucleon spin
decomposition problem with particular emphasis upon the fact that there exists
two physically inequivalent gauge-invariant decompositions (I) and (II)
of the nucleon spin. The difference between these two decompositions
resides in the orbital parts of quarks and gluons, while intrinsic spin
parts of quarks and gluons are just common. The OAMs of quarks and
gluons appearing in the decomposition (I) are the gauge-invariant
dynamical (or mechanical) OAMs, while the OAMs appearing in the
decomposition (II) are the (generalized) canonical OAMs with gauge-invariance.
The key ingredient, which characterizes the difference between these
two OAMs is what-we-call the potential angular momentum.
We clarify the physical meaning of this quantity by using an analogous
but much simpler example from electrodynamics, i.e. a system of
charged particles and photons. It was shown that the potential angular
momentum represents angular momentum associated with the longitudinal
component of the electric field generated by the charged particles.
Remember the fact that the longitudinal component of the electric field
is also the origin of the Coulomb interactions between the charged
particles, although the generation of potential angular momentum needs
the magnetic field as well. Related to the fact that the longitudinal
component of the electric field does not show up in the absence of the charged
particle sources, there arises the ambiguity as to which of charged particles
or the photons the potential angular momentum should be attributed to.
This is essentially the same arbitrariness as which of charged particles
or the photons the Coulomb energy should be attributed to.
If we attribute the potential angular momentum to the charged particle
property, we have an angular momentum decomposition, in which
the angular momentum of the charged particles is given by the (generalized)
canonical OAM. 
On the other hand, if we attribute it to the property of the
photons, the orbital part of the charged particle is given by the
mechanical (or dynamical) OAM. Although the choice is a matter of taste,
it is important to recognize the fact that what is closer to the physical
image of orbital (rotational) motion of charged particles is the mechanical
OAM rather than the canonical OAM, in sharp contradiction to a wide-spread
belief or prepossession. One confirms that the terminology {\it mechanical}
OAM has a legitimate reason for it. This understanding may be of important
physical significance, because, for example, one must recognize clearly which
of dynamical or mechanical OAMs is a relevant quantity, when one tries to
explain the single-spin asymmetry of semi-inclusive hadron productions based
on the orbital angular momenta of nucleon constituents.  

Also addressed in the paper are several other issues left in the
decomposition problem of nucleon spin and momentum.
After verifying the fact that there exist two gauge-invariant decomposition
of the QCD energy-momentum tensor into the quark and gluon contributions,
which are generally nonequivalent, we have verified that the two decompositions
give exactly the same answer as long as the longitudinal momentum sum rule
of the nucleon is concerned. It was further proved that
the two decomposition give the same answer also for the evolution equation
for the momentum fractions of quarks and gluons, which contradicts
Chen et al's claim that the gluons carry much smaller momentum fraction
in the asymptotic limit as compared with the standardly-believed value
of about one-half.

We have also compared the evolution equations of OAMs of quarks and gluons
appearing in the two decompositions of the nucleon spin. We confirmed the
fact that these two types of OAMs obey exactly the same evolution equations
as indicated by the preceding studies. We showed that the
reason of this somewhat mysterious observation can be trace back to the fact
that the potential angular momentum, which gives the difference between the
two types of OAMs, does not contribute to the evolution matrix of
the quark and gluon OAMs. We therefore believe that, through the present
investigation, our understanding about the relation between two different
decomposition of the nucleon spin has been deepened much.

\vspace{0mm}
\begin{acknowledgments}
I would like to thank Elliot Leader and Cedric L\'{o}rce for continual and
many enlightening discussions throughout the study.
This work is supported in part by a Grant-in-Aid for
Scientific Research for Ministry of Education, Culture, Sports, Science
and Technology, Japan (No.~C-21540268)
\end{acknowledgments}

\vspace{4mm}
\appendix

\section{proof that the potential momentum term does not contribute to
the anomalous dimension matrix of quark and gluon momentum fractions.}

The contribution of the potential momentum term to $\gamma^{(2)}_{qq}$
can be obtained by evaluating the matrix element
\begin{equation}
 T_{qq} \ = \ \langle p s \,|\,\int \,d^3 x \,g \,\bar{\psi} (x) \,
 \gamma^+ \,A^+_{phys}(x) \,\psi (x) \,|\,p s \rangle ,
\end{equation}
where $|\,p s \rangle$ is one quark state with momentum $p$ and spin $s$.
The corresponding 1-loop diagram is given by the diagram $(b)$ of Fig.1
except that the vertex $V_B$ is replaced by $V_C$.
Taking care of the Feynman rule explained in the text, we obtain
\begin{eqnarray}
 T_{qq} &=& \int \,\frac{d^4 k}{(2 \,\pi)^4} \,\,\bar{u}(ps) \,\,g \,
 \gamma^+ \,g^{+ \nu} \,t^a \,\frac{i \not\!k}{k^2 + i \,\varepsilon} \,
 ( - \,i \,g \,\gamma^\mu \,t^b) \,u (ps) \nonumber \\
 &\,& \times \ \frac{- \,i \,\delta_{ab}}{(k-p)^2 + i \,\varepsilon} \,\,
 \sum_{\lambda = 1}^2 \,\varepsilon_\mu (k-p,\lambda) \,
 \varepsilon_\nu^* (k-p,\lambda) .
\end{eqnarray}
By using
\begin{equation}
 T_{\mu \nu} \ \equiv \ \sum_{\lambda = 1}^2 \,
 \varepsilon_\mu (k, \lambda) \,\varepsilon_\nu^* (k, \lambda) 
 \ = \ g_{\mu \nu} \ - \ 
 \frac{k_\mu \,n_\nu + k_\nu \,n_\mu}{k \cdot n} ,
\end{equation}
with $n_\mu$ being a light-like vector with $n^2 = 0$ \cite{Wakamatsu11B},
we can write as
\begin{eqnarray}
 T_{qq} &=& - \,g^2 \,C_F \,\int \,\frac{d^4 k}{(2 \,\pi)^4} \,
 \frac{1}{(k^2 + i \,\varepsilon) \,[(k - p)^2 + i \,\varepsilon]} \nonumber \\
 &\,& \times \ \bar{u} (ps) \,\gamma^+ \,g^{+ \nu} \not\!k \,\gamma^\mu \,u (ps) \,
 \left[\,g_{\mu \nu} \ - \ \frac{(k-p)_\mu \,n_\nu + (k-p)_\nu \,n_\mu}
 {(k-p) \cdot n} \,\right].
\end{eqnarray}
Averaging over the spins, we obtain
\begin{eqnarray}
 &\,& \overline{\sum}_{spins} \,\bar{u}(ps) \,\gamma^+ \,g^{+ \nu} 
 \not\!k \,
 \gamma^\mu \,u(ps) \,\,g_{\mu \nu} \ = \ 4 \,k^+ \,p^+, \\
 &\,& \overline{\sum}_{spins} \,\bar{u}(ps) \,\gamma^+ \,g^{+ \nu} 
 \not\!k \,
 \gamma^\mu \,u(ps) \,\,\frac{(k-p)_\mu \,n_\nu + (k-p)_\nu \,n_\mu}
 {(k-p) \cdot n} \ = \ 4 \,k^+ \,p^+ . \ \ \ \ \ \ \ 
\end{eqnarray}
We thus find that the contributions from the two parts of the (modified)
gluon propagator precisely cancel each other, which proves our statement
that the potential momentum term does not contribute to $\gamma^{(2)}_{qq}$.

Next, we consider the following matrix element
\begin{equation}
 T_{qG} \ = \ \langle p \lambda \,|\,\int \,d^3 x \,\,g \,\bar{\psi} \,
 \gamma^+ \,A^+_{phys} \,\psi \,|\,p \lambda \rangle ,
\end{equation}
where $|\,p \lambda \rangle$ is one gluon state with momentum $p$ and polarization
$\lambda$. The 1-loop diagram, that might potentially contribute to this
matrix element, is shown in Fig.3. This gives
\begin{eqnarray}
 T_{qG} &=& \int \,\frac{d^4 k}{(2 \,\pi)^4} \,
 \frac{1}{(k^2 + i \,\varepsilon) \,[(k-p)^2 + i \,\varepsilon]} \nonumber \\
 &\,& \times \ \mbox{Tr} \,[\gamma^+ \,t_a \not\!k \,\gamma^\nu \,t_a \,
 (\not\!k \ - \not\!p) \,] \,\,\varepsilon_+ (p,\lambda) \,
 \varepsilon_\nu^* (p,\lambda) .
\end{eqnarray}
Since the real gluon state has only transverse polarizations, we have
\begin{equation}
 \varepsilon_+ (p, \lambda) \ = \ 0 .
\end{equation}
This ensures that the potential momentum term does not contribute
to the anomalous dimension $\gamma^{(2)}_{qG}$.

\section{Proof that the potential angular momentum term does not contribute
to the evolution matrix for orbital angular momentum.}

We are interested here in the 1-loop contribution to the matrix element
\begin{equation}
 \langle p + \,|\,\hat{L}_{pot} \,|\, p + \rangle ,
\end{equation}
with
\begin{equation}
 \hat{L}_{pot} \ = \ \int \,d^3 x \,\,g \,\bar{\psi} (x) \,\gamma^+ \,
 (x^1 \,A^2_{phys} (x) \ - \ x^2 \,A^1_{phys} (x)) \,\psi (x).
\end{equation}
To avoid singular nature of the matrix element resulting from the explicit
factor of $x^\mu$, it is customary to first consider off-forward matrix
element and to take the forward limit afterwards.
For the off-forward matrix element in quark or gluon state, we have
\begin{eqnarray}
 \langle p^{\prime +} \,|\,\hat{L}_{pot} \,|\, p + \rangle &=& 
 \int \,d^3 x \,\left\{\, x^1 \,\langle p^\prime + \,|\,
 g \,\bar{\psi} (x) \,\gamma^+ \,A^2_{phys} (x) \,\psi (x) \,|\, p + \rangle
 \right. \nonumber \\
 &\,& \hspace{12mm} \left. - \,x^2 \,\langle p^\prime + \,|\,
 g \,\bar{\psi} (x) \,\gamma^+ \,A^1_{phys} (x) \,\psi (x) \,|\, p + \rangle
 \,\right\} \nonumber \\
 &=& (2 \,\pi)^3 \,\left[\,- \,i \,\frac{\partial}{\partial p^\prime_1} \,
 \delta^3 (p^\prime - p) \,\langle p^\prime + \,|\,g \,
 \bar{\psi}(0) \,\gamma^+ \,A^2_{phys} (0) \,\psi (0) \,|\,p + \rangle
 \right. \nonumber \\
 &\,& \hspace{13mm} \left.
 + \,i \,\frac{\partial}{\partial p^\prime_2} \,
 \delta^3 (p^\prime - p) \,\langle p^\prime + \,|\,g \,
 \bar{\psi}(0) \,\gamma^+ \,A^1_{phys} (0) \,\psi (0) \,|\,p + \rangle
 \,\right] . \ \ \ 
\end{eqnarray}
When convoluted with a test function \cite{JTH96}, this gives two terms.
One is
\begin{equation}
 \lim_{p^\prime \rightarrow p} \,
 \left[\,i \,\frac{\partial}{\partial p^\prime_1} \,
 \langle p^\prime + \,|\, g \,\bar{\psi}(0) \,\gamma^+ \,
 A^2_{phys}(0) \, |\,p + \rangle \ - \ i \,
 \frac{\partial}{\partial p^\prime_2} \,
 \langle p^\prime + \,|\, g \,\bar{\psi}(0) \,\gamma^+ \,
 A^1_{phys}(0) \, |\,p + \rangle \,\right],
\end{equation}
which represents the generation of orbital angular momentum from quark
and gluon helicities in the splitting processes. The other is
\begin{equation}
 - \,\frac{1}{p^i} \,\langle p + \,|\,g \,\bar{\psi}(0) \,
 \gamma^+ \,A^i_{phys} (0) \,\psi (0) \,|\,p + \rangle ,
 \hspace{10mm} (i \ : \ \mbox{not summed})
\end{equation}
which represents the self-generation of orbital angular momentum
in the splitting processes.

We first consider the former contribution, which has the structure
\begin{equation}
 \lim_{p^\prime \rightarrow p} \,
 \left(\,i \,\frac{\partial}{\partial p^\prime_1} \,\tilde{T}^2 \ - \ 
 i \,\frac{\partial}{\partial p^\prime_2} \,\tilde{T}^1 \,\right),
\end{equation}
with
\begin{equation}
 \tilde{T}^i \ = \ \langle p^\prime + \,|\,g \,\bar{\psi}(0) \,
 \gamma^+ \,A^i_{phys} (0) \,\psi (0) \,|\,p + \rangle .
\end{equation}
The 1-loop Feynman diagram contributing this matrix element is similar
to that shown in Fig.4. This gives
\begin{eqnarray}
 \tilde{T}^i &=& \frac{1}{2 \,p^+} \,\int \,\frac{d^4 k}{(2 \,\pi)^4} \,
 \bar{u} (p^\prime +) \,g \,\gamma^+ \,g^{i \mu} \,t^a \,
 \frac{i \not\!k}{k^2 + i \,\varepsilon} \,(- \,g \,\gamma^\nu \,t^b) \,
 u (p+) \nonumber \\
 &\,& \hspace{15mm} \times \ 
 \frac{- \,i \,\delta_{ab}}{(k-p)^2 + i \,\varepsilon} \,
 \sum_{\lambda = 1}^2 \,\varepsilon_\mu (k-p,\lambda) \,
 \varepsilon_\nu^* (k-p,\lambda) .
\end{eqnarray}
From this, we obtain
\begin{eqnarray}
 i \,\frac{\partial}{\partial p^\prime +} \,\tilde{T}^2 &=&
 \frac{g^2 \,C_F}{2 \,p^+} \,\int \,\frac{d^4 k}{(2 \,\pi)^4} \,
 \frac{\partial}{\partial p^\prime_1} \,\bar{u} (p^\prime_1) \,
 \gamma^+ \not\!k \,\gamma^\nu \,u (p +) \nonumber \\
 &\,& \times \ \frac{1}{(k^2 + i \,\varepsilon) \,
 [(k-p)^2 + i \,\varepsilon]} \,\left[\,\delta^i_\nu \ - \ 
 \frac{(k-p)^i \,n_\nu}{(k-p) \cdot n} \,\right] ,
\end{eqnarray}
where we have used the fact that $n^i = 0$ for $i = 1,2$.
Using the explicit form of the light-cone spinors
\begin{equation}
 u(p +) \ = \ \frac{1}{\sqrt{2}\,\sqrt{\sqrt{2}\,p^+}} \,
 \left( \begin{array}{c}
 \sqrt{2} \,p^+ \\
 0 \\
 \sqrt{2} \,p^+ \\
 0 \\
 \end{array} \right), \hspace{10mm}
 u(p^\prime +) \ = \ \frac{1}{\sqrt{2}\,\sqrt{\sqrt{2}\,p^+}} \,
 \left( \begin{array}{c}
 \sqrt{2} \,p^+ \\
 p^\prime_1 + i \,p^\prime_2 \\
 \sqrt{2} \,p^+ \\
 p^\prime_1 + i \,p^\prime_2 \\
 \end{array} \right),
\end{equation}
it can be shown that
\begin{eqnarray}
 &\,& \lim_{p^\prime \rightarrow p} \,\frac{\partial}{\partial p^\prime_1} \,
 \bar{u} (p^\prime) \,\gamma^+ \not\!k \,\gamma^\nu \,u (p) \,\delta^i_\nu
 \ = \ 0, \\
 &\,& \lim_{p^\prime \rightarrow p} \,\frac{\partial}{\partial p^\prime_1} \,
 \bar{u} (p^\prime) \,\gamma^+ \not\!k \,\gamma^\nu \,u (p) \,
 (k - p)^i \,n_\nu \ = \ 0.
\end{eqnarray}
Then, we find that
\begin{equation}
 \lim_{p^\prime \rightarrow p} \,i \,\frac{\partial}{\partial p^\prime_1} \,
 \tilde{T}^2 \ = \ 0 .
\end{equation}
Similarly
\begin{equation}
 \lim_{p^\prime \rightarrow p} \,i \,\frac{\partial}{\partial p^\prime_2} \,
 \tilde{T}^1 \ = \ 0 ,
\end{equation}
In this way, we find that
\begin{equation}
 \lim_{p^\prime \rightarrow p} \,
 \left(\, i \,\frac{\partial}{\partial p^\prime_1} \,\tilde{T}^2 \ - \ 
 i \,\frac{\partial}{\partial p^\prime_2} \,\tilde{T}^1 \,\right) \ = \ 0.
\end{equation}

Next we consider the term corresponding to self-generation of the orbital
angular momentum in the splitting processes, which takes of the form : 
\begin{equation}
 - \,\frac{1}{p^i} \,\langle p + \,|\,g \,\bar{\psi}(0) \,\gamma^+ \,
 A^i_{phys} (0) \,\psi (0) \,|\,p + \rangle \ = \ - \,\frac{1}{p^i} \,T^i .
\end{equation}
We find that
\begin{eqnarray}
 T^i &=& \frac{1}{2 \,p^+} \,\int \,\frac{d^4 k}{(2 \,\pi)^4} \,
 \bar{u} (p+) \,g \,\gamma^+ \,g^{i \mu} \,t^a \,
 \frac{i \,\not\!k}{k^2 + i \,\varepsilon} \,
 (- \,g \,\gamma^\nu \,t^a) \,u (p +) \nonumber \\
 &\,& \times \ \frac{- \,i \,\delta_{ab}}{(k-p)^2 + i \,\varepsilon} \,
 \left[\, g_{\mu \nu} \ - \ \frac{(k-p)_\mu \,n_\nu + n_\mu \,(k-p)_\nu}
 {(k-p) \cdot n} \,\right] \nonumber \\
 &=& - \,\frac{i \,g^2 \,C_F}{2 \,p^+} \,
 \int \,\frac{d^4 k}{(2 \,\pi)^4} \,\bar{u} (p+) \,\gamma^+ 
 \not\!k \,\gamma^\nu \,u (p+) \nonumber \\
 &\,& \hspace{5mm} \times \ 
 \frac{1}{(k^2 + i \,\varepsilon) \,[(k-p)^2 + i \,\varepsilon} \,
 \left[\, \delta^i_\nu \ - \ \frac{(k-p)^i \,n_\nu}{(k-p) \cdot n} \,\right]
\end{eqnarray}
Using the relation
\begin{equation}
 u (p+) \,\gamma^+ \not\!k \,\gamma^\nu \,u (p+) \,
 \left[\, \delta^i_\nu \ - \ \frac{(k-p)^i \,n_\nu}{(k-p) \cdot n} \,\right]
 \ = \ 
 2 \,(k^+ \,p^i \ - \ p^+ \,k^i) \,\,\frac{k^+ + p^+}{k^+ - p^+} ,
\end{equation}
we have 
\begin{eqnarray}
 T^i &=& - \,\frac{2 \,i \,g^2 \,C_F}{2 \,p^+} \,
 \int \,\frac{d^4 k}{(2 \,\pi)^4} \,
 \frac{1}{(k^2 + i \,\varepsilon) \,[(k-p)^2 + i \,\varepsilon]}
 \,\, (k^+ \,p^i \ - \ p^+ \,k^i) \,\,\frac{k^+ + p^+}{k^+ - p^+}
 \ \ \ \  . 
\end{eqnarray}
Carrying out $k^-$ integration, shifting the variable $\bm{k}_\perp$ tp
$\bm{k}^\prime_\perp = \bm{k}_\perp - x \,\bm{p}_\perp$ and trading $k^+$
for $x \,p^+$, we obtain
\begin{equation}
 T^i \ = \ \frac{g^2 \,C_F}{2 \,(2 \,\pi)^3} \,
 \int \,\frac{d^2 \bm{k}^\prime_\perp}{\bm{k}^{\prime 2}_\perp} \,
 \int_0^1 \,(x \ - \ x) \,\,\frac{x + 1}{x - 1} \ = \ 0.
\end{equation}
This means that the contributions from the two parts of the modified
gluon propagator cancel each other. We therefore confirm the fact that
the potential angular momentum term, which distinguishes the two
decompositions of the nucleon spin, does not contribute to the
evolution matrix for the quark OAM. We emphasize that our proof here
is not bound to the choice of gauge.

\vspace{10mm}

\end{document}